\documentclass[12pt]{article}
\usepackage{a4wide}
\usepackage{latexsym}
\usepackage{amsmath}
\usepackage{amsfonts}
\usepackage{amscd}
\usepackage{cite}
\usepackage{slashed} 
\usepackage{pslatex}
\usepackage{graphicx}
\usepackage[latin1,utf8]{inputenc}
\usepackage[T1]{fontenc}

\allowdisplaybreaks

\newcommand{\bq}{\begin{eqnarray}}
\newcommand{\eq}{\end{eqnarray}}
\newcommand{\eps}{\varepsilon}

\newcommand{\intprod}{\mathbin{\raisebox{\depth}{\scalebox{1}[-1]{$\lnot$}}}}

\usepackage{amssymb}
\usepackage{amsthm}
\usepackage{mathtools}
\usepackage{enumerate}
\usepackage{hyperref}
\usepackage[all]{xy}
\usepackage{adjustbox}
\usepackage{color}
\usepackage{lscape}

\usepackage{tikz}
\usepackage{tcolorbox}
\usepackage[thinlines]{easytable} 

\theoremstyle{plain}
\newtheorem{theorem}{Theorem}[section]
\newtheorem{corollary}[theorem]{Corollary}

\theoremstyle{definition}

\newtheorem{definition}[theorem]{Definition}
\newtheorem{example}[theorem]{Example}

\newtheorem{remark}[theorem]{Remark}

\begin{document}

\thispagestyle{empty}

\begin{flushright}
  MITP/18-091
\end{flushright}

\vspace{1.5cm}

\begin{center}
  {\Large\bf Rationalizing roots: an algorithmic approach \\
  }
  \vspace{1cm}
  {\large Marco Besier ${}^{a,b}$, Duco van Straten ${}^{a}$ and Stefan Weinzierl ${}^{b}$\\
  \vspace{1cm}
      {\small \em ${}^{a}$ Institut f{\"u}r Mathematik, }\\
      {\small \em Johannes Gutenberg-Universit{\"a}t Mainz,}\\
      {\small \em D - 55099 Mainz, Germany}\\
  \vspace{0.5cm}
      {\small \em ${}^{b}$ PRISMA Cluster of Excellence, Institut f{\"u}r Physik, }\\
      {\small \em Johannes Gutenberg-Universit{\"a}t Mainz,}\\
      {\small \em D - 55099 Mainz, Germany}\\
  } 
\end{center}

\vspace{2cm}

\begin{abstract}
  {
In the computation of Feynman integrals which evaluate to multiple polylogarithms
  one encounters quite often square roots.
  To express the Feynman integral in terms of multiple polylogarithms,
  one seeks a transformation of variables, which rationalizes the square roots.
  In this paper, we give an algorithm for rationalizing roots.
  The algorithm is applicable whenever the algebraic hypersurface associated with the root has a point of multiplicity $(d-1)$, where $d$ is the degree of the algebraic hypersurface.
  We show that one can use the algorithm iteratively to rationalize multiple roots simultaneously.
  Several examples from high energy physics are discussed.
   }
\end{abstract}

\vspace*{\fill}

\newpage

\section{Introduction}
\label{sec:introduction}

We may view a Feynman integral as a function of the number of space-time dimensions $D$ and the kinematic variables 
(Lorentz scalar products $p_i \cdot p_j$ and masses $m_j$).
Within dimensional regularization, we are interested in the Laurent expansion in $\eps=(4-D)/2$.
Each coefficient of this Laurent expansion is then a function of the kinematic variables.
 
A significant number of Feynman integrals evaluate to multiple polylogarithms, 
meaning that each term of the Laurent expansion in $\eps$ may be expressed as a linear combination of multiple polylogarithms with
prefactors being algebraic functions of the kinematic variables.
The arguments of the multiple polylogarithms are again functions of the kinematic variables.
These arguments are often called letters, and the set of all letters the symbol alphabet.
Multiple polylogarithms are a special case of iterated integrals, where all integration kernels are of the form 
\bq
\label{standard_integration_kernel}
 \omega_j & = & \frac{dy}{y-z_j},
\eq
where the $z_j$'s are independent of the integration variable $y$ (but may depend on the kinematic variables).
In Feynman integral computations it is not uncommon to encounter integration kernels
like
\bq 
 \frac{dy}{\sqrt{\left(y-z_1\right)\left(y-z_2\right)}}.
\eq
These are not of the form as in equation~(\ref{standard_integration_kernel}).
To express the result in terms of multiple polylogarithms, we seek
a transformation of the integration variable which transforms the integration kernel
into a rational function.
Subsequently using partial fraction decomposition, we may express the integral
in terms of the integration kernels as in equation~(\ref{standard_integration_kernel})
plus trivial integrations.

In this paper, we consider the problem of finding a rational parametrization for the integration kernels.
For several specific examples a rational parametrization is known \cite{Becchetti:2017abb,Broadhurst:1993mw,Fleischer:1998nb,Aglietti:2004tq,Gehrmann:2018yef,Henn:2013woa,Lee:2017oca}.
For Feynman integrals in massless theories and enjoying a dual conformal symmetry the use of momentum twistor variables
has been advocated quite recently \cite{Bourjaily:2018aeq}.
The use of these variables automatically rationalizes a subset of the occurring letters.
There is also an interesting connection of the symbol alphabet in massless theories with cluster $A$-coordinates \cite{Golden:2014xqa,Golden:2013xva}.
An alternative to iterated integrals are nested sums \cite{Moch:2001zr}.
In this approach roots enter through binomial or inverse binomial sums \cite{Davydychev:2003mv,Weinzierl:2004bn,Kalmykov:2006hu,Ablinger:2014bra}.

Despite the considerations in the appendix of \cite{Becchetti:2017abb}, however, to the best of our knowledge, no systematic approach for finding a rational parametrization and working in the massless and the massive case alike has been put forward in the physics community.
Such a systematic approach is the topic of this paper.
We show that the problem can be tackled with methods from elementary algebraic geometry \cite{MR3330490, Lemmermeyer:2011, Perez:2005, Sendra:2008}.
We present an algorithm -- well-known from algebraic geometry -- that rationalizes the given root by first associating an algebraic hypersurface to the root and then parametrizing this hypersurface by an $n$-parameter family of lines, where $n$ is the number of variables occurring in the root. 
This method can be used, whenever the associated hypersurface is irreducible and has at least one point of multiplicity $(d-1)$, where $d$ is the degree of the defining polynomial of the hypersurface. In particular, the algorithm is neither constrained by the number of variables $n$ occurring in the root nor by the degree $d$ of the associated hypersurface. 
Using the method iteratively, we are also able to find parametrizations that rationalize multiple roots simultaneously.
We show that for a variety of roots appearing in physics, which are known to be rationalizable, we can produce rational parametrizations with the help of our method. 
In some cases, we are even able to optimize known parametrizations.\\
However, not all algebraic hypersurfaces have a rational parametrization, the simplest counter-example being given by a 
non-degenerate elliptic curve.
These curves occur in Feynman integral computations.
The corresponding Feynman integrals do not evaluate to multiple polylogarithms.\\
As already mentioned, our algorithm works if the algebraic hypersurface associated with the root is irreducible and possesses a point of multiplicity $(d-1)$.
Let us stress that the fact that an algebraic hypersurface does not have a point of multiplicity $(d-1)$ does not
imply that we cannot parametrize the hypersurface by rational functions.
The hypersurface may or may not have a rational parametrization.
There exist more advanced methods from algebraic geometry which may be used in such a case, and we will provide a simple example in appendix \ref{sec:limitationsOfTheAlgorithm}.
However, weighing the required mathematical overload for these advanced methods against the fact that we can produce parametrizations for a large class of rationalizable roots occurring in physics with a basic algorithm, we find it useful and efficient to restrict in this paper our attention to roots whose associated algebraic hypersurface can be parametrized by lines.\\
This paper is organized as follows: 
in section \ref{sec:motivation} we give a motivational example to illustrate the need for rational parametrizations in the context of loop calculations.
In section \ref{sec:warmup} we present a short warm-up exercise to get a first idea on how to find rational parametrizations systematically.
In section \ref{sec:mathematicalPreliminaries} we introduce the required mathematical framework.
Section \ref{sec:theAlgorithm} and \ref{sec:applicationInPhysics} represent the central part of the paper: here we formulate the rationalization algorithm and show how the method is to be used by discussing several examples related to physics.
In section \ref{sec:summary} we draw our conclusions. 
Also, we include an appendix where we present some examples of non-rationalizable roots such as roots associated with elliptic curves and K3 surfaces and show how one can prove the non-rationalizability of a given root in a mathematically rigorous way.
Furthermore, we formulate and prove a theorem that can, for a particular type of roots, be used to make our algorithm even more efficient by reducing the degree of the associated hypersurface from $d$ to $\frac{d}{2}+1$ for $d$ even. 
This theorem is, for instance, applicable to the Gramian roots of \cite{Bourjaily:2018aeq}.


\section{Motivation}
\label{sec:motivation}

\begin{figure}[!htbp]
\begin{center}
\includegraphics[scale=1.8]{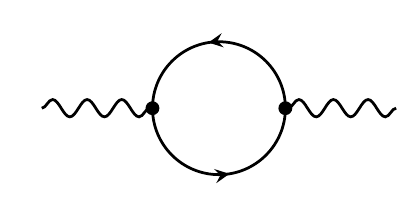}
\end{center}
\caption{
\it A one-loop Feynman diagram contributing to the gauge boson self-energy with a massive fermion loop.
}
\label{fig_one_loop_self_energy}
\end{figure}

Let us consider a simple example, where a square root occurs.
Figure~\ref{fig_one_loop_self_energy} shows a one-loop Feynman diagram relevant to the self-energy of a
gauge boson. 
Let us assume that the fermion circulating in the loop has mass $m$.
We denote the external momentum of the gauge boson by $p$
and set $x:=p^2/m^2$.
We work in dimensional regularization and we denote the dimension of space-time by $D=4-2\eps$.
Associated with the diagram of figure~\ref{fig_one_loop_self_energy} is the set of Feynman integrals
\bq
\label{def_oneloopfamily}
 I_{\nu_1 \nu_2}\left( D, x \right)
 & = &
 \left(m^2\right)^{\nu_1+\nu_2-\frac{D}{2}}
 \int \frac{d^Dk}{i \pi^{\frac{D}{2}}}
 \frac{1}{\left[m^2-k^2 \vphantom{\left(k-p\right)^2}\right]^{\nu_1} \left[m^2-\left(k-p\right)^2\right]^{\nu_2}}.
\eq
Here, $\nu_1$ and $\nu_2$ denote two integers.
Integration-by-parts identities \cite{Chetyrkin:1981qh,Tkachov:1981wb}
relate integrals with different indices $(\nu_1,\nu_2)$.
For this example, we may express any integral $I_{\nu_1 \nu_2}$ as a linear combination of two master integrals.
As the two master integrals we may take
\bq
\label{def_precanonical}
 I_1\left(\eps,x\right)
 \; = \;
 I_{20}\left(4-2\eps,x\right),
 & &
 I_2\left(\eps,x\right)
 \; = \;
 I_{21}\left(4-2\eps,x\right).
\eq
The differential equation for $\vec{I}=(I_1,I_2)^T$ reads
\bq
 \frac{d}{dx} \vec{I} 
 & = &
 \left( \begin{array}{cc}
 0 & 0 \\
 \frac{\eps}{4 x} - \frac{\eps}{4\left(x-4\right)} & -\frac{1}{2x} - \frac{1+2\eps}{2\left(x-4\right)} \\
 \end{array} \right)
 \vec{I}.
\eq
It is desirable to bring the differential equation into a form, where the only explicit $\eps$-dependence is through a prefactor $\eps$ on the 
right-hand side \cite{Henn:2013pwa}.
This can be achieved by changing the basis of master integrals. 
We divide $I_2$ by its maximal cut.
The maximal cut of $I_2$ is given by
\bq
 \mathrm{MaxCut}\; I_2 
 & \sim &
 \frac{1}{\sqrt{-x\left(4-x\right)}},
\eq
up to a constant prefactor.
We set
\bq
\label{def_canonical_basis}
 J_1\left(\eps,x\right)
 \; = \;
 2 \eps I_1,
 & &
 J_2\left(\eps,x\right)
 \; = \;
 2 \eps \sqrt{-x\left(4-x\right)} I_2.
\eq
In the basis $\vec{J}=(J_1,J_2)^T$ the differential equation reads
\bq
 \frac{d}{dx} \vec{J} 
 & = &
 \eps
 \left( \begin{array}{cc}
 0 & 0 \\
 - \frac{1}{\sqrt{-x\left(4-x\right)}} & - \frac{1}{x-4} \\
 \end{array} \right)
 \vec{J}.
\eq
This differential equation is now nicely in $\eps$-form. However, a square root $\sqrt{-x(4-x)}$ sneaked in.
We want to rationalize the square root by a change of variables from $x$ to a new variable $t$.
For the case at hand a solution is well-known \cite{Fleischer:1998nb}:
setting 
\bq
 x & = & - \frac{\left(1-t\right)^2}{t}
\eq
removes the square root.
Indeed, we have
\bq
 \frac{d}{dt} \vec{J} 
 & = &
 \eps
 \left( \begin{array}{cc}
 0 & 0 \\
 - \frac{1}{t} & \frac{1}{t} - \frac{2}{t+1} \\
 \end{array} \right)
 \vec{J}.
\eq
This differential equation is now easily solved in terms of harmonic polylogarithms.

\section{Warm-up exercise}
\label{sec:warmup}

Let us suppose we encounter the square root $\sqrt{1-x^2}$ in our physical problem at hand and we need to find an appropriate transformation $\varphi_x: t\mapsto \varphi_x(t)$ that turns
\begin{equation}
    \sqrt{1-\left(\varphi_x(t)\right)^2} 
\end{equation}
into a rational function of $t$. One easily checks that the parametrization
\begin{equation}
    \varphi_x(t)=\frac{1-t^2}{1+t^2}
\end{equation}
solves the problem, leading to 
\begin{equation}
    \sqrt{1-\left(\varphi_x(t)\right)^2}=\frac{2t}{1+t^2}.
\end{equation}
Now, the interesting observation is that we can construct the expression for $\varphi_x(t)$ in a systematic way. First of all, we name the square root by defining $y:=\sqrt{1-x^2}$. Taking the square, we observe that this yields the defining equation
\begin{equation}
    x^2+y^2-1=0
\end{equation}
of the unit circle. Thus, it is quite natural to say that the square root $\sqrt{1-x^2}$ \textit{is associated with} the unit circle.\\
The circle is one representative of a fundamental class of mathematical objects. It defines an algebraic curve, which in turn is a special case of the more general concept of \textit{algebraic hypersurfaces}, which are defined to be the set of zeros of a polynomial. We see that asking for a rational change of variables $\varphi_x(t)$ which rationalizes the square root $y=\sqrt{1-x^2}$ is the same as asking for rational functions $(\varphi_x(t),\varphi_y(t))$ which parametrize the unit circle. If one can find such rational functions, one would call the circle a \textit{rational} algebraic hypersurface.\\
For the square root $\sqrt{1-x^2}$, the solution to the rationalization problem is known since antiquity: consider a fixed point $P$ on the circle and a variable point $Q$ moving on a line not passing through $P$. Then look at the second point of intersection $R$ of the line $PQ$ with the circle. We observe that, if $Q$ traces its line, then $R$ traces the circle. If we take the point $P$ to be $(-1,0)$ and assume $Q$ to move along the $y$-axis, i.e., $Q=(0,t)$, then the defining equation of the line $PQ$ is given by $y=t(1+x)$ from which we find the parametrization
\begin{equation}
    R(t):=(\varphi_x(t),\varphi_y(t))=\left(\frac{1-t^2}{1+t^2},\frac{2t}{1+t^2}\right)
\label{eq:GeomParamCircle}
\end{equation}
of the unit circle by a short calculation: simply determine the intersection points of the line $PQ:y=t(1+x)$ and the circle $x^2+y^2=1$. The first point of intersection is $P$, the second one yields $R(t)$.\\
Remarkably, this parametrization was already known 1500 BC \cite{ifrah1999universal}.
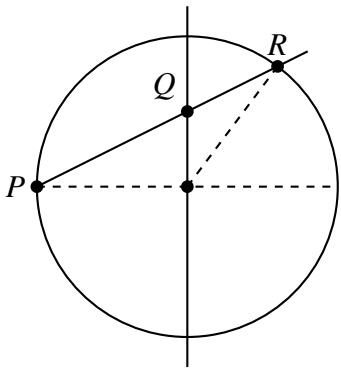
\begin{figure}[!htbp]
    \centering
    \begin{tikzpicture}[thick, scale=2]
    \draw[dashed] (-1,0) -- (1,0);
    \draw[dashed] (0,0) -- (0.6,0.8);
    \draw (-1,0) -- (0.8,0.9);
    \draw (0,-1.2) -- (0,1.2);
    \filldraw[black] (0,0) circle (1pt) node[anchor=west] {};
    \filldraw[black] (0,0.5) circle (1pt) node[anchor=south east] {$Q$};
    \filldraw[black] (0.6,0.8) circle (1pt) node[anchor=south] {$R$};
    \filldraw[black] (-1,0) circle (1pt) node[anchor=east] {$P$};
    \draw (0,0) circle (1);
    \end{tikzpicture}  
    \caption{\it Parametrizing the circle by a 1-parameter family of lines.}
    \label{fig:my_label}
\end{figure}
\begin{remark}
    Notice that, to calculate the expression for $R(t)$, one solely needs rational operations (addition, subtraction, multiplication, division) on polynomial expressions with coefficients in $\mathbb{Q}$. This is precisely the reason why the above method returns a \textit{rational function} of $t$.\\
    We ensure rational \textit{coefficients} by choosing $P$ to be a point with all coordinate entries lying in $\mathbb{Q}$. In principle, nothing prevents us from taking $P\notin \mathbb{Q}^2$, e.g., choosing $P=\left(-\frac{1}{\sqrt{2}},-\frac{1}{\sqrt{2}}\right)$ as the starting point of our construction. Still, the method would return a rational function. However, the coefficients of this rational function would no longer be rational, but rather contain factors of $\sqrt{2}$ (cf. example \ref{ex:sqrt2Param}). 
\label{rem:reasonWhyWarmupWorks}
\end{remark}
Despite the simplicity of this example, we do not want to perform a geometric construction for every root we need to rationalize. Instead, we are looking for an algebraic, algorithmic approach to the problem. In the sequel of this paper, we will show that such an algorithm can indeed be found for a large class of roots.

\section{Mathematical preliminaries}
\label{sec:mathematicalPreliminaries}

In this section, we will introduce the mathematical framework, which is needed to study the rationalization problems we encounter in physics. We begin by introducing the notion of affine algebraic hypersurfaces.
\begin{definition}
     An \textit{affine algebraic hypersurface} of dimension $(n-1)$ over $\mathbb{C}$ is a set 
    \begin{equation}
        V=\left\{(a_1,\ldots,a_n)\in\mathbb{A}^{n}(\mathbb{C})\hspace{2pt}\vert\hspace{2pt}f(a_1,\ldots,a_n)=0\right\},
    \end{equation}
    where $f(x_1,\ldots,x_n)\in \mathbb{C}[x_1,\ldots,x_n]$ is a non-constant polynomial in $n$ variables and $\mathbb{A}^{n}(\mathbb{C})\equiv\mathbb{C}^n$ is the affine space of dimension $n$ over the complex numbers. We call $f$ the \textit{defining polynomial} of $V$. If deg$(f)=d$, then $d$ is called the \textit{degree} of $V$, denoted by deg$(V)$. If $f=\prod_{i=1}^{m}f_{i}^{k_i}$, where $m,k_1,\ldots,k_m\in \mathbb{N}$ and $f_{i}$ are the irreducible factors of $f$, we say that the hypersurface defined by each polynomial $f_{i}$, is a \textit{component} of $V$. Moreover, the hypersurface $V$ is said to be \textit{irreducible} if its defining polynomial is irreducible, i.e., $m=k_1=1$. 
    Notice that the defining polynomial of $V$ is unique only up to multiplication by non-zero constants $c\in \mathbb{C}$ and powers $k_i\in \mathbb{N}$ of the factors of $f$:
    e.g., the polynomial $g=cf^2$ defines the same hypersurface as $f$.
    We will therefore define algebraic hypersurfaces via reduced polynomials, i.e., $c=k_1=\ldots=k_m=1$.
    Notice that one should not confuse reduced polynomials with polynomials that are irreducible:
    for instance, the polynomial $f(x,y)=x^2-y^2=(x+y)\cdot(x-y)$ is reduced but it is not irreducible over $\mathbb{C}$.
\end{definition}
Let us have a look at some prominent examples of hypersurfaces:
for $n=2$ one obtains a plane affine algebraic \textit{curve} of degree $d$. 
Curves of degree 1 are called \textit{lines}, of degree 2 \textit{conics}, of degree 3 \textit{cubics}, etc. 
For example, the unit circle with defining polynomial $f(x,y)=x^2+y^2-1$ is a conic, whereas the elliptic curve defined by $f(x,y)=y^2-x^3-x-1$ is a cubic.
For $n=3$ we obtain affine algebraic \textit{surfaces} of degree $d$. 
If, for instance, $n=3$ and $d=4$, we speak of an \textit{affine quartic surface}.\\
The unit circle is an example of a \textit{smooth} curve. 
However, one often encounters singular points on hypersurfaces, and we will soon see that precisely these points, especially the ones with very high multiplicity, are crucial for our rationalization method to work out. 
Therefore, let us define these notions properly.\\
The \textit{tangent space} $T_pV$ of a hypersurface $V: f(x_1,\ldots,x_n)=0$ at a point $p=(p_1,\ldots,p_n)$ is itself an algebraic hypersurface and given by
\begin{equation}
    T_pV: \sum_{i=1}^{n}\frac{\partial f}{\partial x_i}(p)\cdot(x_i-p_i)=0.
\end{equation}
The singular points of a hypersurface are precisely the points which do not allow for a well-defined tangent space.
\begin{definition}
    If $V$ is an affine algebraic hypersurface of dimension $(n-1)$ over $\mathbb{C}$ with defining equation $f(x_1,\ldots,x_n)=0$, then a point $p\in \mathbb{A}^n(\mathbb{C})$ that satisfies
    \begin{equation}
        f(p)=\frac{\partial f}{\partial x_1}(p)=\cdots=\frac{\partial f}{\partial x_n}(p)=0
        \label{eq:SingularityCondition}
    \end{equation}
    is called a \textit{singular point} of $V$. All non-singular points of $V$ are called \textit{regular} points.
\end{definition}
\begin{example}
    Consider the nodal cubic $V: f(x,y)=y^2-x^3-x^2=0$ and the point $p=(0,0)$. One easily verifies $f(p)=\frac{\partial f}{\partial x}(p)=\frac{\partial f}{\partial y}(p)=0$, showing that this curve has a singular point at the origin. 
    \begin{figure}[!htbp]
    \centering
    \includegraphics[scale=0.1]{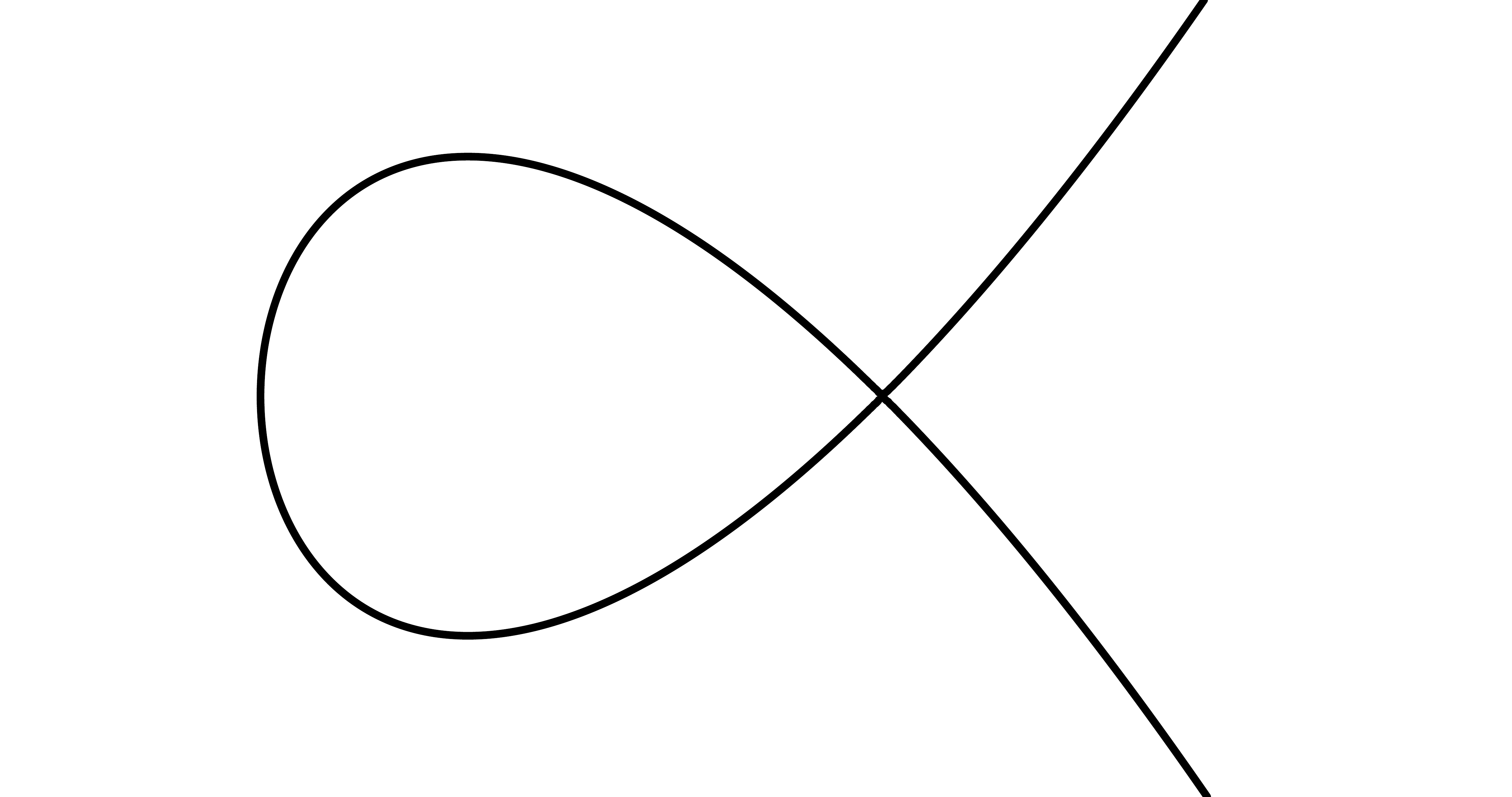}
    \caption{\textit{The nodal cubic $V: y^2-x^3-x^2=0$.}}
    \label{fig:nodalcubic}
    \end{figure}
\label{ex:nodalcubic}
\end{example}
\begin{definition}
    A point $p\in V$ of a hypersurface $V\subset\mathbb{A}^{n}(\mathbb{C})$ with defining polynomial $f$ is said to be \textit{of multiplicity} $r\in \mathbb{N}_0$, if there exists at least one non-vanishing $r$-th partial derivative
\begin{equation}
    \frac{\partial^{i_1+\cdots+i_n} f}{\partial x_1^{i_1}\cdots \partial x_n^{i_n}}(p)\neq0,\hspace{4pt}\text{where }i_1+\cdots+i_n=r
\end{equation}
and, at the same time, all partial derivatives of lower order vanish at $p$, i.e.,
\begin{equation}
    \frac{\partial^{i_1+\cdots+i_n} f}{\partial x_1^{i_1}\cdots \partial x_n^{i_n}}(p)=0\hspace{4pt}\text{with }i_1+\cdots+i_n=k\hspace{4pt}\text{for all }k=1,\ldots,r-1. 
\end{equation}
We write mult$_p(V)=r$. Points $p\notin V$, which do not belong to the hypersurface $V$, are of multiplicity 0. The \textit{regular} points of $V$ are of multiplicity 1. Every point $p\in V$ with mult$_p(V)>1$ is, inevitably, a singular point of $V$. In the case of curves, the multiplicity of a point is, loosely speaking, the number of smooth branches passing through this point.
\label{def:Multiplicity}
\end{definition}
Looking at example \ref{ex:nodalcubic} again, we see that the nodal cubic has a point of multiplicity 2 at the origin. 
All other points $p\in V$ with $p\neq(0,0)$ are regular points with mult$_p(V)=1$.
\begin{remark}
    The notion of multiplicity is invariant under linear changes of coordinates. Therefore, we may alternatively define the multiplicity of a point as follows: note that we can always write
\begin{equation}
    f(x_1,\ldots,x_n)=f_0(x_1,\ldots,x_n)+\cdots+f_d(x_1,\ldots,x_n),
\end{equation} 
where $f(x_1,\ldots,x_n)$ is a polynomial of degree $d$, $f_k(x_1,\ldots,x_n)$ with $k=0,\ldots,d$ are homogeneous polynomials of degree $k$ (cf. definition \ref{def:projectivehypersurfaces}) and $f_d(x_1,\ldots,x_n)$ is non-zero. 
    We call $f_k(x_1,\ldots,x_n)$ the \textit{homogenous components} of $f(x_1,\ldots,x_n)$.
    The multiplicity of an affine algebraic hypersurface $V: f(x_1,\ldots,x_n)=0$ at the origin of $\mathbb{A}^n(\mathbb{C})$ can also be defined to be the minimum of the degrees of the non-zero homogeneous components of $f$. 
\label{rem:multAtOriginAndMultInvariant}
\end{remark}
So, by taking into account remark \ref{rem:multAtOriginAndMultInvariant}, we can also determine the multiplicity of a point $p\in V\subset\mathbb{A}^n(\mathbb{C})$ by moving $p$ to the origin via a linear change of coordinates and reading off the minimum of the degrees of the non-zero homogeneous components of $f$.
Let us conclude these considerations with the following important corollary:
\begin{corollary}
    Whenever we encounter a hypersurface $V:f(x_1,\ldots,x_n)=0$ of degree $d$ with a point $p=(p_1,\ldots,p_n)$ of multiplicity $r$ and move $p=(p_1,\ldots,p_n)$ to the origin by considering
\begin{equation}
    g(x_1,\ldots,x_n):=f(x_1+p_1,\ldots,x_n+p_n),
\end{equation}
then $g(x_1,\ldots,x_n)$ can always be written as
\begin{equation}
    g(x_1,\ldots,x_n)=g_r(x_1,\ldots,x_n)+\cdots+g_d(x_1,\ldots,x_n),
\end{equation}
where $g_k(x_1,\ldots,x_n)$ with $k=r,\ldots,d$ are the homogeneous components of $g(x_1,\ldots,x_n)$.
\label{cor:multAtOrigin}
\end{corollary}
As we will see in the upcoming section, solely working with hypersurfaces in affine space will not be sufficient. Some information about the hypersurface is hidden when working in the affine framework. Luckily, it turns out that by passing to the projective closure one picks up this hidden information. Therefore, it will be useful to consider hypersurfaces in projective space.
\begin{definition}
    A \textit{projective algebraic hypersurface} of dimension $(n-1)$ and degree $d$ over $\mathbb{C}$ is defined as the set 
    \begin{equation}
        \tilde{V}=\{[a_1:\ldots:a_{n+1}]\in\mathbb{P}^{n}(\mathbb{C})\hspace{2pt}\vert\hspace{2pt}F(a_1,\ldots,a_{n+1})=0\},
    \end{equation}
    where $F(x_1,\ldots,x_{n+1})\in \mathbb{C}[x_1,\ldots,x_{n+1}]$ is a non-constant polynomial in $(n+1)$ variables, which is homogeneous of degree $d$, i.e., for $\lambda\in \mathbb{C}$ non-zero, one has
\begin{equation}
    F(\lambda x_1,\ldots,\lambda x_{n+1})=\lambda^d F(x_1,\ldots,x_{n+1}).
\end{equation}
$\mathbb{P}^{n}(\mathbb{C})$ is the $n$-dimensional complex projective space and $[a_1:\ldots:a_{n+1}]$ are points of $\mathbb{P}^{n}(\mathbb{C})$, denoted in homogeneous coordinates.
\label{def:projectivehypersurfaces}
\end{definition}
All notions introduced for affine hypersurfaces, in particular the notions of singular points and multiplicities, carry over to the projective setting in the natural way.
Now, the point is that we can turn every affine algebraic hypersurface $V$ of degree $d$ with defining equation $f(x_1,\ldots,x_n)=0$ into a corresponding projective hypersurface $\tilde{V}$ of degree $d$ via homogenization of $f$. This means that $\tilde{V}$ is defined by the polynomial
\begin{equation}
    f^h(x_1,\ldots,x_n,z):=z^{\text{deg}(f)}f\left(\frac{x_1}{z},\ldots,\frac{x_n}{z}\right),
\end{equation}
where $f^h(x_1,\ldots,x_n,z)$ is homogeneous of degree $d$, by definition. $\tilde{V}$ is called the \textit{projective closure} of $V$.\\
On the other hand, if $F(x_1,\ldots,x_n,z)$ is the defining polynomial of a projective hypersurface, then $F(x_1,\ldots,x_n,1)$ is the defining polynomial of an affine hypersurface, which consists of the points of the projective hypersurface whose last coordinate is non-zero. These two procedures are, in fact, reciprocal to one another in the sense that, as $f^h(x_1,\ldots,x_n,1)=f(x_1,\ldots,x_n)$ and, if $f$ is defined by $f(x_1,\ldots,x_n):=F(x_1,\ldots,x_n,1)$, then $f^h(x_1,\ldots,x_n,z)=F(x_1,\ldots,x_n,z)$, as soon as the homogeneous polynomial $F$ is not divisible by $z$.\\
One may think of the affine hypersurface as one \textit{chart} of the corresponding projective hypersurface. 
The points of the projective closure of an affine hypersurface which do not belong to the affine part are called \textit{points at infinity}. 
So, if the affine hypersurface $V$ is defined by the polynomial $F(x_1,\ldots,x_n,1)$ and is consequently a chart of the projective hypersurface $\tilde{V}: F(x_1,\ldots,x_n,z)=0$, then the points at infinity of the affine chart $V$ would be the points $[a_1:\ldots:a_n:0]\in \tilde{V}$, i.e., the points $[a_1:\ldots:a_n:0]\in \mathbb{P}^n(\mathbb{C})$ fulfilling $F(a_1,\ldots,a_n,0)=0$.\\
In total, a projective hypersurface $\tilde{V}\subset \mathbb{P}^{n}(\mathbb{C})$ of dimension $(n-1)$ has $(n+1)$ affine charts, each obtained by setting one of the $(n+1)$ projective coordinates equal to 1, leaving the others as affine coordinates of the respective chart.\\
To have an easy example in mind, let us consider the unit circle again, i.e., the affine hypersurface $V_{xy}: x^2+y^2-1=0$. The projective closure of $V_{xy}$ is the projective hypersurface $\tilde{V}: x^2+y^2-z^2=0$. Consequently, the unit circle is one affine chart of $\tilde{V}$. However, the two hyperbolas $V_{xz}: x^2+1-z^2=0$ and $V_{yz}: 1+y^2-z^2=0$ are affine charts of $\tilde{V}$, as well. So, from a projective point of view, these two hyperbolas and the unit circle are the same curve.\\
Furthermore, we observe that each of the hyperbolas $V_{xz}$ and $V_{yz}$ has one point at infinity, namely $[1:0:1]$ and $[0:1:1]$, respectively, whereas the unit circle $V_{xy}$ has the two complex points $[1:i:0]$ and $[1:-i:0]$ at infinity.

\section{The algorithm}
\label{sec:theAlgorithm}

Let us now introduce the class of roots we will be able to rationalize in a straightforward, algorithmic manner. We will provide a more precise definition as soon as we have studied some first examples (cf. definition \ref{def:perfectroots}). A given root is called \textit{perfect} if the associated algebraic hypersurface of degree $d$ is irreducible and has a point of multiplicity $(d-1)$.\\
Now, why is this notion useful? We have seen that we had to use a quite manual geometric construction (cf. section \ref{sec:warmup}) to rationalize the circle. In practice, however, we do not want to draw curves or surfaces to find rational parametrizations. We rather need an easy algebraic algorithm which allows us to rationalize the given root straightforwardly. The point is, now, that for perfect roots, we find such an algorithm. To get an idea of how it works, let us consider $\sqrt{1-x^2}$ again.
\begin{example}
    As we have already seen before, the hypersurface associated with $y:=\sqrt{1-x^2}$ is the unit circle $V: f(x,y)=x^2+y^2-1=0$. Since $d=\text{deg}(V)=2$, finding a point of multiplicity $(d-1)$ is very easy. Recalling that \textit{regular} points $p\in V$ are of multiplicity 1, we may choose any fixed regular point $p_0$ on the circle, e.g., $p_0=(-1,0)\in V$ as in section \ref{sec:warmup}. Notice that the irreducibility of $f$ together with the existence of such a point shows that $\sqrt{1-x^2}$ is a perfect root in the above sense.\\
    The next step is to translate $p_0$ to the origin, i.e., we send $x\mapsto x+1$ and $y\mapsto y$. The polynomial $f$ becomes
\begin{equation}
    f(x,y)=f_1(x,y)+f_2(x,y)
\end{equation}
with homogeneous components
\begin{equation}
    f_1(x,y)=-2x\hspace{4pt}\text{and}\hspace{4pt}f_2(x,y)=x^2+y^2
\end{equation}
of degree 1 and 2, respectively.\\
As in section \ref{sec:warmup}, let us consider a family of lines $y=tx$ through $p_0$. We determine the two intersection points of each of the lines with the circle $V: f(x,y)=0$ by plugging the line equation into $f(x,y)=0$. We obtain
\begin{equation}
    0=f_1(x,tx)+f_2(x,tx)=xf_1(1,t)+x^2f_2(1,t).
\label{eq:intersectingLineAndCircle}
\end{equation}
The solution $x=0$ gives $p_0$. The second solution yields
\begin{equation}
    x=-\frac{f_1(1,t)}{f_2(1,t)},\hspace{4pt}y=-t\frac{f_1(1,t)}{f_2(1,t)}.
\end{equation}
Switching back to the original setting by translating $x\mapsto x-1$ and $y\mapsto y$, we see that
\begin{equation}
    \varphi_x(t)=-\frac{f_1(1,t)}{f_2(1,t)}-1,\hspace{4pt}\varphi_y(t)=-t\frac{f_1(1,t)}{f_2(1,t)}
\end{equation}
is precisely parametrization (\ref{eq:GeomParamCircle}), which we already encountered in section \ref{sec:warmup}. In particular, $\varphi_x(t)$ provides the sought after change of variables that rationalizes $\sqrt{1-x^2}$.
\label{ex:AlgebraicCircleParam}
\end{example}
\begin{remark}
Notice that we had a choice in picking the family of lines through $p_0$ which we intersected with the circle in equation (\ref{eq:intersectingLineAndCircle}). In fact, one can easily produce a different rational parametrization by considering a different family of lines. For instance, take the family to be $x=ty$, substitute $x$ with $ty$ in equation (\ref{eq:intersectingLineAndCircle}) and solve for $y$.
\label{rem:differentFamiliesInCircleExample}
\end{remark}
\begin{example}
Consider the square root $y:=\sqrt{x^3+x^2}$. The associated hypersurface $V$ is irreducible, of degree $d=3$ and given by the nodal cubic $V: f(x,y)=y^2-x^3-x^2=0$, which we have already discussed in example \ref{ex:nodalcubic}. We have seen that $V$ has a point of multiplicity $3-1=2$ at the origin, proving $\sqrt{x^3+x^2}$ to be a perfect root. Analogously to example \ref{ex:AlgebraicCircleParam}, we can parametrize $V$ by the family of lines $y=tx$, yielding
\begin{equation}
    \varphi_x(t)=t^2-1,\hspace{4pt}\varphi_y(t)=t^3-t.
\end{equation}
However, if we intersect $V$ with the family $x=ty$ instead, we obtain a different parametrization:
\begin{equation}
    \phi_x(t)=\frac{1-t^2}{t^2},\hspace{4pt}\phi_y(t)=\frac{1-t^2}{t^3}.
\end{equation}
\label{ex:nodalCubicParam}
\end{example}
In the case of example \ref{ex:nodalCubicParam}, there is only a single point of multiplicity $(d-1)$, namely the origin. For the unit circle $V: f(x,y)=x^2+y^2-1=0$, however, \textit{every single point of} $V$ has multiplicity $(d-1)$. This allows for the construction of even more rational parametrizations of the circle, by simply choosing another starting point $p_0$. Notice that this also works for ellipses, hyperbolas and, in fact, any irreducible conic.
\begin{example}
    Consider the circle $V: f(x,y)=x^2+y^2-1=0$ and choose $p_0=\left(-\frac{1}{\sqrt{2}},-\frac{1}{\sqrt{2}}\right)$ as opposed to our former choice $(-1,0)$. Moving $p_0$ to the origin, intersecting the circle with the family of lines $y=tx$ and finally translating $p_0$ back to its original position, we obtain the following parametrization:
\begin{equation}
    \varphi_x(t)=\frac{1-(t-2)t}{\sqrt{2}\left(1+t^2\right)},\hspace{4pt}\varphi_y(t)=\frac{t^2+2t-1}{\sqrt{2}\left(1+t^2\right)}.
\end{equation}
    Not only does this show that we get different parametrizations for different choices of $p_0$, but it also makes the statements of remark \ref{rem:reasonWhyWarmupWorks} more explicit: choosing a point with irrational coordinates still leads to a rational parametrization. The coefficients of these rational functions will, however, no longer be rational, but will involve the irrationalities of $p_0$.\\
    Irrational coefficients are not particularly desired, so we try to avoid them in the lucky situation of having multiple points of multiplicity $(d-1)$ to choose from. So, whenever possible, we choose our starting point $p_0$ to have rational coordinates.
\label{ex:sqrt2Param}
\end{example}
Before we move on to a more sophisticated example of a rationalization procedure, let us demonstrate how to identify the hypersurface associated with a given root.
The easiest case is the one, where we face a square root
\begin{equation}
    \sqrt{\frac{p(x_1,\ldots,x_n)}{q(x_1,\ldots,x_n)}}
\label{eq:simplestAssociatedhypersurface}
\end{equation}
of a rational function, where $p(x_1,\ldots,x_n)$, $q(x_1,\ldots,x_n)\in \mathbb{C}[x_1,\ldots,x_n]$ are multivariate non-zero polynomials. The associated hypersurface is simply obtained by naming the root, e.g., denote it by $u$, squaring the resulting equation and clearing the denominator. Thus, the hypersurface $V$ associated with the root (\ref{eq:simplestAssociatedhypersurface}) is given as
\begin{equation}
    V: f(u,x_1,\ldots,x_n)=q(x_1,\ldots,x_n)u^2-p(x_1,\ldots,x_n)=0.
\end{equation}
Of course, square roots of rational functions are the standard use case, i.e., they are the most likely to appear in physical contexts. However, we want to point out that the method presented here is \textit{not} restricted to simple square roots but can also be applied to roots of higher order and, by suitable exponentiation and clearing of denominators, even to arbitrary algebraic functions. In particular, this allows for the study of nested roots, as well. In this way, we find an associated hypersurface for any algebraic function. For example, the hypersurface $V$ associated with the 3-rd root
\begin{equation}
    \sqrt[\leftroot{0}\uproot{0}3]{x^3+x^2}
\label{eq:perfectThirdRoot}
\end{equation}
is given by
\begin{equation}
    V: f(u,x)=u^3-x^3-x^2=0.
\end{equation}
Indeed, the root (\ref{eq:perfectThirdRoot}) is perfect, since $f$ is irreducible, deg$(V)=3$ and $V$ has a point of multiplicity 2 at the origin. 
We may also consider rather involved nested roots, for instance
\begin{equation}
    \sqrt{\frac{-2x^2-3x+\sqrt{16x^3+9x^2}}{2}}.
\label{eq:3leavedCloverRoot}
\end{equation}
Again, let us call this root $u$. By squaring, clearing the denominator and squaring once more, we obtain
\begin{equation}
    \left(2u^2+2x^2+3x\right)^2=16x^3+9x^2.
\end{equation}
Simplifying this equation, we see that the hypersurface $V$ associated with (\ref{eq:3leavedCloverRoot}) is given by
\begin{equation}
    V: f(u,x)=x^4-x^3+2x^2u^2+3xu^2+u^4=0.
\end{equation}
We observe deg$(V)=4$ and, taking into account remark \ref{rem:multAtOriginAndMultInvariant}, we immediately see that $V$ has a point of multiplicity 3 at the origin. Furthermore, $f$ is irreducible. Consequently, the root (\ref{eq:3leavedCloverRoot}) is perfect and in turn rationalizable by the method presented in this paper.\\
Before formulating the full algorithm, let us investigate one example, which is a bit more sophisticated since it contains more than one variable.
\begin{example}
    Suppose we want to rationalize the square root
\begin{equation}
    \sqrt{\frac{x^4+4x^2y^2+4}{4x^2}}.
\end{equation}
Denoting this root by $u$, its associated affine hypersurface is irreducible and given by
\begin{equation}
    V: f(u,x,y)=x^4+4x^2y^2+4-4u^2x^2=0.
\end{equation}
First of all, we want to know whether or not $u$ is a perfect root. We have deg$(V)=4$. Thus, we need to prove the existence of some $p_0\in V$ with $\text{mult}_{p_0}(V)=4-1=3$. So, in order to find $p_0$, we first try to solve condition (\ref{eq:SingularityCondition}), since a point of multiplicity $(d-1)$ has, inevitably, to be a \textit{singular} point of the hypersurface as soon as $d>2$. Unfortunately, it turns out that the hypersurface $V$ does not possess any affine singularities. In particular, it does not possess any affine point of multiplicity $3$.\\
However, there is one last hope: there might be points at infinity, which we are not able to see in the affine setting, and some of these points might be singular with the required multiplicity.\\
Indeed, one finds that the projective closure
\begin{equation}
    \tilde{V}: F(u,x,y,z)=x^4+4x^2y^2+4z^4-4u^2x^2=0 
\end{equation}
does have a singular point at $p_0=[u_0:x_0:y_0:z_0]=[1:0:1:0]$ and, recalling definition \ref{def:Multiplicity}, one can check $\text{mult}_{p_0}(\tilde{V})=3$. So, although not immediately obvious, the square root $u$ is perfect.\\
To actually rationalize $u$, we can just look at the problem from another point of view, i.e., from another affine chart $V^\prime$ of $\tilde{V}$, for which $p_0$ is not at infinity. For instance, we may consider the chart in which the first homogeneous coordinate is equal to $1$ via the map
\begin{align}
\begin{split}
        \mathbb{P}^3(\mathbb{C})&\rightarrow\mathbb{A}^3(\mathbb{C})\\
        [u:x:y:z]&\mapsto\left(\frac{x}{u},\frac{y}{u},\frac{z}{u}\right)=:(x^\prime,y^\prime,z^\prime),\hspace{8pt}u\neq0.
\end{split}
\label{eq:ProjectiveCoordinateTrafo}
\end{align}
The singularity $p_0\in\tilde{V}\subset \mathbb{P}^3(\mathbb{C})$ is mapped to the affine point $p_0^\prime:=(0,1,0)\in V^\prime\subset \mathbb{A}^3(\mathbb{C})$.
So, finally, we have a quite similar situation as in example \ref{ex:AlgebraicCircleParam} and are now able to apply the same strategy.\\
Defining $\bar{f}\left(x^\prime,y^\prime,z^\prime\right):=F\left(1,x^\prime,y^\prime,z^\prime\right)$, we translate $p_0^\prime$ to the origin of the affine chart we are working in, i.e., we map $x^\prime\mapsto x^\prime$, $y^\prime\mapsto y^\prime-1$ and $z^\prime\mapsto z^\prime$. This yields
\begin{equation}
    \bar{f}\left(x^\prime,y^\prime,z^\prime\right)=\bar{f}_3\left(x^\prime,y^\prime,z^\prime\right)+\bar{f}_4\left(x^\prime,y^\prime,z^\prime\right)
\end{equation}
with homogeneous components
\begin{equation}
    \bar{f}_3\left(x^\prime,y^\prime,z^\prime\right)=8\left(x^\prime\right)^2y^\prime\hspace{4pt}\text{and}\hspace{4pt}\bar{f}_4\left(x^\prime,y^\prime,z^\prime\right)=\left(x^\prime\right)^4+4\left(z^\prime\right)^4+4\left(x^\prime\right)^2\left(y^\prime\right)^2
\end{equation}
of degree 3 and 4, respectively.\\
In contrast to example \ref{ex:AlgebraicCircleParam}, we now have three instead of only two affine variables involved so that we now need to consider a 2-parameter family of lines passing through $p_0^\prime$. Practically, this means we set $y^\prime=t_1x^\prime$ and $z^\prime=t_2x^\prime$ and plug these into the equation $\bar{f}\left(x^\prime,y^\prime,z^\prime\right)=0$. We obtain
\begin{equation}
0=\bar{f}_3\left(x^\prime,t_1x^\prime,t_2x^\prime\right)+\bar{f}_4\left(x^\prime,t_1x^\prime,t_2x^\prime\right)=\left(x^\prime\right)^3\bar{f}_3\left(1,t_1,t_2\right)+\left(x^\prime\right)^4\bar{f}_4\left(1,t_1,t_2\right).
\end{equation}
The solution $x^\prime=0$ gives $p_0^\prime$, whereas the second solution yields
\begin{align}
\begin{split}
    x^\prime&=-\frac{\bar{f}_3(1,t_1,t_2)}{\bar{f}_4(1,t_1,t_2)}\\
    y^\prime&=-t_1\frac{\bar{f}_3(1,t_1,t_2)}{\bar{f}_4(1,t_1,t_2)}\\
    z^\prime&=-t_2\frac{\bar{f}_3(1,t_1,t_2)}{\bar{f}_4(1,t_1,t_2)}.
\end{split}
\end{align}
Translating back via $x^\prime\mapsto x^\prime$, $y^\prime\mapsto y^\prime+1$ and $z^\prime\mapsto z^\prime$, we obtain
\begin{align}
\begin{split}
     x^\prime&=-\frac{8t_1}{4t_2^4+4t_1^2+1}\\
    y^\prime&=-\frac{8t_1^2}{4t_2^4+4t_1^2+1}+1\\
    z^\prime&=-\frac{8t_1t_2}{4t_2^4+4t_1^2+1}.   
\end{split}
\label{eq:VPrimeParamExample2}
\end{align}
Recall that we are still working in the affine chart $V^\prime$. So, the last remaining step is to switch back to the original affine chart $V$, which is the actual hypersurface associated with the square root $u$. This is simply done by substituting
\begin{equation}
    x^\prime=\frac{x}{u},\hspace{4pt}y^\prime=\frac{y}{u}\hspace{4pt}\text{and}\hspace{4pt}z^\prime=\frac{z}{u} 
\label{eq:InvertingProjectiveTrafo}
\end{equation}
in (\ref{eq:VPrimeParamExample2}) and solving for $x,y$ and $u$ while putting $z=1$. Finally, we obtain a parametrization of $V$:
\begin{align}
\begin{split}
    \varphi_u(t_1,t_2)&=-\frac{4t_2^4+4t_1^2+1}{8t_1t_2}\\
    \varphi_x(t_1,t_2)&=\frac{1}{t_2}\\
    \varphi_y(t_1,t_2)&=-\frac{4t_2^4-4t_1^2+1}{8t_1t_2}.
\end{split}
\end{align}
In particular, $\varphi_x(t_1,t_2)$ and $\varphi_y(t_1,t_2)$ provide the sought after change of variables, which rationalizes the original square root $u$.
\label{ex:InvolvedExample}
\end{example}
Having discussed these examples, we are now ready to give the precise definition of perfect roots and formulate the general algorithm to rationalize these.
\begin{definition}
    An algebraic function $q(x_1,\ldots,x_n)$ in $n$ variables is called a \textit{perfect root}, if the projective closure $\tilde{V}$ of its associated affine hypersurface $V$ is irreducible and has at least one point $p_0\in \tilde{V}$ with $\text{mult}_{p_0}(\tilde{V})=d-1$, where $d=\text{deg}(V)=\text{deg}(\tilde{V})$.
\label{def:perfectroots}
\end{definition}
For perfect roots, there is the following parametrization algorithm:
\newpage

\begin{tcolorbox}
\textbf{Algorithm:}
\begin{itemize}
\item[] \textbf{Input:} A perfect root $u:=q(x_1,\ldots,x_n)$.
\item[] \textbf{Output:} A rational parametrization of the hypersurface $V: f(u,x_1,\ldots,x_n)=0$ of degree $d$ associated with the root, which, in particular, provides the sought after change of variables that rationalizes the given root.
\end{itemize}
\begin{itemize}
\item[1)] Determine a point $p_0\in V$ with $\text{mult}_{p_0}(V)=d-1$. If $d=2$, one may take any regular point on $V$.
\item[2)] If $p_0$ is not at infinity, continue with step 3) and finish with step 4).\\
 In case $p_0$ is at infinity, consider, via a coordinate change like (\ref{eq:ProjectiveCoordinateTrafo}) (cf. example \ref{ex:InvolvedExample}), another affine chart $V^\prime$ of the projective closure $\tilde{V}$ of $V$ in which $p_0$ is not at infinity and continue with steps 3) and 4) and finish with step 5).
\item[3)] With $p_0=(a_0,a_1,\ldots,a_n)$, compute $g(u,x_1,\ldots,x_n):=f(u+a_0,x_1+a_1,\ldots,x_n+a_n)$ and write (cf. corollary \ref{cor:multAtOrigin})
\begin{equation}
    g(u,x_1,\ldots,x_n)=g_{d-1}(u,x_1,\ldots,x_n)+g_{d}(u,x_1,\ldots,x_n),
\end{equation}
where $g_{d-1}$ and $g_{d}$ are homogeneous components of degree $(d-1)$ and $d$.
\item[4)] Return
\begin{align}
\begin{split}
        \varphi_u(t_1,\ldots,t_n)&=-\frac{g_{d-1}(1,t_1,\ldots,t_n)}{g_{d}(1,t_1,\ldots,t_n)}+a_0\\
    \varphi_{x_1}(t_1,\ldots,t_n)&=-t_1\frac{g_{d-1}(1,t_1,\ldots,t_n)}{g_{d}(1,t_1,\ldots,t_n)}+a_1\\
    &\vdots\\
    \varphi_{x_n}(t_1,\ldots,t_n)&=-t_n\frac{g_{d-1}(1,t_1,\ldots,t_n)}{g_{d}(1,t_1,\ldots,t_n)}+a_n.
\end{split}
\label{eq:ALgorithmOutput}
\end{align}
\item[5)] To (\ref{eq:ALgorithmOutput}), apply the \textit{inverse} of the change of coordinates that was used to switch from $V$ to $V^\prime$.
Refer to (\ref{eq:InvertingProjectiveTrafo}) in example \ref{ex:InvolvedExample} to see how this works in detail.
\end{itemize}
\end{tcolorbox}
\newpage

At this point, it is expedient to give some remarks:
step 4) of the algorithm returns the rational parametrization for $V$. Therefore,
\begin{equation}
    \varphi_{x_1}(t_1,\ldots,t_n),\ldots,\varphi_{x_n}(t_1,\ldots,t_n)
\end{equation}
give a change of variables, which rationalizes the perfect root $u$.\\
The reason why this algorithm works out is due to a generalization of B\'{e}zout's theorem \cite{MR2215193}:
by choosing a point of multiplicity $(d-1)$ as the starting point of the family of lines that we use to parametrize the hypersurface associated with the root, we make sure that a generic line of the family has only one unique further point of intersection with the hypersurface.
For instance, if we would choose a point other than the origin in the case of the nodal cubic (cf. example \ref{ex:nodalcubic} and \ref{ex:nodalCubicParam}) and consider a family of lines through this point, then each line would, in general, have \textit{two} further intersection points with the curve.\\
Notice that our analysis is not restricted to a single root at a time.
In fact, we can apply the above algorithm iteratively to construct a \textit{single} parametrization that rationalizes \textit{multiple} roots simultaneously (cf. example \ref{ex:rationalizingMultipleRoots}).
\begin{remark}
    As we have already seen in remark \ref{rem:differentFamiliesInCircleExample} and example \ref{ex:nodalCubicParam}, we can easily obtain other parametrizations of the perfect root by changing the $n$-parameter family of lines that we use to parametrize its associated hypersurface. Thus, in addition to (\ref{eq:ALgorithmOutput}), we may also return
\begin{align}
    \begin{split}
    \varphi_u(t_1,\ldots,t_n)&=-t_1\frac{g_{d-1}(t_1,1,t_2,\ldots,t_n)}{g_{d}(t_1,1,t_2,\ldots,t_n)}+a_0\\
    \varphi_{x_1}(t_1,\ldots,t_n)&=-\frac{g_{d-1}(t_1,1,t_2,\ldots,t_n)}{g_{d}(t_1,1,t_2,\ldots,t_n)}+a_1\\
    \varphi_{x_2}(t_1,\ldots,t_n)&=-t_2\frac{g_{d-1}(t_1,1,t_2,\ldots,t_n)}{g_{d}(t_1,1,t_2,\ldots,t_n)}+a_2\\
    &\vdots\\
    \varphi_{x_n}(t_1,\ldots,t_n)&=-t_n\frac{g_{d-1}(t_1,1,t_2,\ldots,t_n)}{g_{d}(t_1,1,t_2,\ldots,t_n)}+a_n\\
    &\vdots\\
    &\vdots\\
    \varphi_u(t_1,\ldots,t_n)&=-t_1\frac{g_{d-1}(t_1,\ldots,t_n,1)}{g_{d}(t_1,\ldots,t_n,1)}+a_0\\
    \varphi_{x_{1}}(t_1,\ldots,t_n)&=-t_2\frac{g_{d-1}(t_1,\ldots,t_n,1)}{g_{d}(t_1,\ldots,t_n,1)}+a_{1}\\
    &\vdots\\
    \varphi_{x_{n-1}}(t_1,\ldots,t_n)&=-t_n\frac{g_{d-1}(t_1,\ldots,t_n,1)}{g_{d}(t_1,\ldots,t_n,1)}+a_{n-1}\\
    \varphi_{x_n}(t_1,\ldots,t_n)&=-\frac{g_{d-1}(t_1,\ldots,t_n,1)}{g_{d}(t_1,\ldots,t_n,1)}+a_n
\end{split}
\end{align}
and use the most convenient of these parametrizations to rationalize the root under consideration.
\label{rem:otherOutputsOfTheAlgorithm}
\end{remark}
The parametrizations obtained by choosing a different point of high multiplicity or different families of lines are not the only parametrizations one can construct.
In fact, given a parametrization
\begin{equation}
    \varphi_{u}(t_1,\ldots,t_n),\varphi_{x_1}(t_1,\ldots,t_n),\ldots,\varphi_{x_n}(t_1,\ldots,t_n)
\end{equation}
of $V$ and $R_1(t_1),\ldots,R_n(t_n)$ are, for instance, arbitrary non-constant rational functions, then 
\begin{equation}
    \varphi_{u}(R_1(t_1),\ldots,R_n(t_n)),\varphi_{x_1}(R_1(t_1),\ldots,R_n(t_n)),\ldots,\varphi_{x_n}(R_1(t_1),\ldots,R_n(t_n))
\end{equation}
parametrizes $V$, as well.

\section{Applications in physics}
\label{sec:applicationInPhysics}

Let us now justify the effort we made in the previous chapters by considering rationalization problems of some quite recent papers \cite{Becchetti:2017abb, Bourjaily:2018aeq, Caron-Huot:2016owq} that are directly related to high energy physics.
\begin{example}
The first physical example we will present is the square root $\Delta_6^{\{123456\}}$ of \cite{Bourjaily:2018aeq} which also appears in the hexagon function bootstrap in planar maximally supersymmetric Yang-Mills theory \cite{Caron-Huot:2016owq}. It is given by
\begin{equation}
    \Delta_6^{\{123456\}}=\sqrt{\left(1-u_1-u_2-u_3\right)^2-4u_1u_2u_3}.
\label{eq:Delta6}
\end{equation}
Setting $u:=\Delta_6^{\{123456\}}$, we see that the defining polynomial $f$ of the hypersurface $V$ associated with this root is
\begin{equation}
    f(u,u_1,u_2,u_3)=\left(1-u_1-u_2-u_3\right)^2-4u_1u_2u_3-u^2.
\end{equation}
Since $\text{deg}(V)=3$, we are looking for a point $p_0\in V$ with $\text{mult}_{p_0}(V)=3-1=2$. We find four points that satisfy this condition:
\begin{equation}
    \{(0,0,0,1),(0,0,1,0),(0,1,0,0),(0,1,1,1)\}.
\end{equation}
So, for instance, we may choose $p_0=(0,0,0,1)$. Next, we consider the polynomial
\begin{align}
\begin{split}
       g(u,u_1,u_2,u_3):&=f(u+0,u_1+0,u_2+0,u_3+1)\\
&=\left(1-u_1-u_2-(u_3+1)\right)^2-4u_1u_2(u_3+1)-u^2\\
&=g_2(u,u_1,u_2,u_3)+g_3(u,u_1,u_2,u_3) 
\end{split}
\end{align}
with
\begin{align}
\begin{split}
  g_2(u,u_1,u_2,u_3)&=(u_1+u_2+u_3)^2-4u_1u_2-u^2\\
  g_3(u,u_1,u_2,u_3)&=-4u_1u_2u_3.
\end{split}
\end{align}
Using our algorithm, a rational parametrization of $V$ is readily computed to be
\begin{align}
\begin{split}
        \varphi_u(t_1,t_2,t_3)&=-\frac{g_2(1,t_1,t_2,t_3)}{g_3(1,t_1,t_2,t_3)}\\
&=\frac{(t_1+t_2+t_3)^2-4t_1t_2-1}{4t_1t_2t_3}\\
    \varphi_{u_1}(t_1,t_2,t_3)&=-t_1\frac{g_2(1,t_1,t_2,t_3)}{g_3(1,t_1,t_2,t_3)}\\
&=\frac{(t_1+t_2+t_3)^2-4t_1t_2-1}{4t_2t_3}\\
    \varphi_{u_2}(t_1,t_2,t_3)&=-t_2\frac{g_2(1,t_1,t_2,t_3)}{g_3(1,t_1,t_2,t_3)}\\
&=\frac{(t_1+t_2+t_3)^2-4t_1t_2-1}{4t_1t_3}\\
    \varphi_{u_3}(t_1,t_2,t_3)&=-t_3\frac{g_2(1,t_1,t_2,t_3)}{g_3(1,t_1,t_2,t_3)}+1\\
&=\frac{(t_1+t_2+t_3)^2-4t_1t_2-1}{4t_1t_2}+1.
\end{split}
\end{align}
\end{example}
\begin{example}
The second example we want to study is relevant in the context of planar QCD
massive corrections to di-photon and di-jet
hadro-production and was first solved in \cite{Becchetti:2017abb}. Consider the following set of square roots:
\begin{equation}
    \mathcal{A}:=\left\{\sqrt{u+1},\sqrt{u-1},\sqrt{v+1},\sqrt{u+v+1}\right\}.
\label{eq:SetOfSimultaneousRoots}
\end{equation}
Compared to (\ref{eq:Delta6}), the individual roots of (\ref{eq:SetOfSimultaneousRoots}) look rather simple. The difficulty is, however, that we need to determine a change of variables that rationalizes \textit{all} square roots of the set (\ref{eq:SetOfSimultaneousRoots}) \textit{simultaneously}.
\par\vspace{\baselineskip}
\textit{Rationalizing }$\sqrt{u+1}$.
\par\vspace{\baselineskip}

Denoting the square root by $w$, the hypersurface $V$ associated with $\sqrt{u+1}$ is the conic curve defined by the polynomial $f(u,w)=w^2-u-1$. Since $d=2$, any regular point $p_0$ of $V$ fulfills $\text{mult}_{p_0}(V)=d-1$. For instance, we may choose $p_0=(u_0,w_0)=(-1,0)$. We define the polynomial
\begin{equation}
    g(u,w):=f(u-1,w+0)=g_1(u,w)+g_2(u,w)
\end{equation}
with
\begin{equation}
    g_1(u,w)=-u\hspace{4pt}\text{and}\hspace{4pt}g_2(u,w)=w^2.
\end{equation}
The rationalization algorithm yields the following parametrization:
\begin{align}
\begin{split}
        \varphi_u(t_1)&=-\frac{g_1(1,t_1)}{g_2(1,t_1)}-1=\frac{1-t_1^2}{t_1^2}\\
    \varphi_w(t_1)&=-t_1\frac{g_1(1,t_1)}{g_2(1,t_1)}=\frac{1}{t_1}. 
\end{split}
\end{align}
\textit{Rationalizing }$\sqrt{u-1}$.
\par\vspace{\baselineskip}
Next, we want to find a parametrization which rationalizes the square root $\sqrt{u-1}$. However, at the same time, we have to guarantee that this new parametrization $\tilde{\varphi}_u(t_2)$ preserves the property of rationalizing the first square root $\sqrt{u+1}$. In order to achieve this, the first step is to substitute the expression for $\varphi_u(t_1)$ in the square root $\sqrt{u-1}$:
\begin{equation}
    \sqrt{\varphi_u(t_1)-1}=\sqrt{\frac{1-2t_1^2}{t_1^2}}.
\end{equation}
Since the denominator of the right-hand side already is a perfect square, the root we actually need to rationalize is the one in the numerator. Denoting this root by $q$, the associated hypersurface $W$ is the conic curve defined by the polynomial $f(t_1,q)=q^2+2t_1^2-1$. Again, since $d=2$, any regular point $p_0$ of $W$ fulfills $\text{mult}_{p_0}(W)=d-1$. For instance, we may choose $p_0=(t_{1_0},q_0)=(0,1)$. We define the polynomial
\begin{equation}
    g(t_1,q):=f(t_1+0,q+1)=g_1(t_1,q)+g_2(t_1,q)
\end{equation}
with
\begin{equation}
    g_1(t_1,q)=2q\hspace{4pt}\text{and}\hspace{4pt}g_2(t_1,q)=q^2+2t_1^2.
\end{equation}
The rationalization algorithm yields:
\begin{align}
\begin{split}
        \varphi_{t_1}(t_2)&=-\frac{g_1(1,t_2)}{g_2(1,t_2)}=-\frac{2t_2}{t_2^2+2}\\
    \varphi_q(t_2)&=-t_2\frac{g_1(1,t_2)}{g_2(1,t_2)}+1=1-\frac{2t_2^2}{t_2^2+2}. 
\end{split}
\end{align}
Now, we can write down the coordinate change $\tilde{\varphi}_u(t_2)$ for $u$ which rationalizes the two square roots $\sqrt{u-1}$ and $\sqrt{u+1}$ simultaneously:
\begin{equation}
        \tilde{\varphi}_u(t_2):=\varphi_u(\varphi_{t_1}(t_2))=\frac{t_2^4+4}{4t_2^2}.
\label{eq:uPlusMinusOneRootsParam}
\end{equation}
Indeed, we can check that, plugging $\tilde{\varphi}_u(t_2)$ into the two square roots, we obtain rational expressions
\begin{align}
\begin{split}
        \sqrt{\tilde{\varphi}_u(t_2)+1}&=\frac{t_2^2+2}{2t_2}\\
\sqrt{\tilde{\varphi}_u(t_2)-1}&=\frac{t_2^2-2}{2t_2}.
\end{split}
\end{align}
Notice that, compared to the change of variables
\begin{equation}
    \tilde{\varphi}_u(t_2)=\frac{\left(t_2^2+1\right)(t_2(t_2+4)+5)}{4(t_2+1)^2}
\end{equation} 
given in \cite{Becchetti:2017abb}, the result we obtained in (\ref{eq:uPlusMinusOneRootsParam}) is more compact and can thus be considered a slight improvement over the known parametrization.
\par\vspace{\baselineskip}
\textit{Rationalizing }$\sqrt{v+1}$.
\par\vspace{\baselineskip}
In principle, this could be done in the exact same manner as $\sqrt{u+1}$. A change of variables which rationalizes $\sqrt{v+1}$ is thus given by
\begin{equation}
    \tilde{\varphi}_v(t_v)=\frac{1-t_v^2}{t_v^2}.
\label{eq:unappropriateVParam}
\end{equation}
However, as we will see below, working with (\ref{eq:unappropriateVParam}) will yield a sextic surface when we try to rationalize $\sqrt{u+v+1}$ in the upcoming iteration step. Although this surface does have a rational parametrization, we are usually looking for the lowest degree possible. In fact, a more appropriate change of variables to rationalize $\sqrt{v+1}$ is provided by simply using
\begin{equation}
    \tilde{\varphi}_v(t_v)=t_v^2-1,
\label{eq:paramForVRoot}
\end{equation}
which is the second, alternative output we get from the rationalization algorithm (cf. remark \ref{rem:otherOutputsOfTheAlgorithm}). We will see that, using (\ref{eq:paramForVRoot}), the next iteration step will only lead to a quartic instead of a sextic surface.
\par\vspace{\baselineskip}
\textit{Rationalizing }$\sqrt{u+v+1}$.
\par\vspace{\baselineskip}
Let us now rationalize the last square root of the set (\ref{eq:SetOfSimultaneousRoots}). Assuming that the first three square roots of $\mathcal{A}$ are already rationalized by means of the transformation
\begin{align}
\begin{split}
        \tilde{\varphi}_u(t_u)&=\frac{t_u^4+4}{4t_u^2}\\
\tilde{\varphi}_v(t_v)&=t_v^2-1,
\end{split}
\label{eq:final3RootsParam}
\end{align}
we can express the remaining square root $p:=\sqrt{u+v+1}$ in terms of the variables $t_u$ and $t_v$:
\begin{equation}
    p=\sqrt{\frac{t_u^4+4t_u^2t_v^2+4}{4t_u^2}},
\label{eq:1+u+vRoot}
\end{equation}
The relevant algebraic hypersurface for which we need to find a rational parametrization is a quartic surface defined by the polynomial equation
\begin{equation}
    0=t_u^4+4t_v^2t_u^2+4-4p^2t_u^2.
\end{equation}
The attentive reader may have noticed that the root (\ref{eq:1+u+vRoot}) is precisely the square root we have already studied in example \ref{ex:InvolvedExample}. A parametrization of (\ref{eq:1+u+vRoot}) is therefore given by
\begin{align}
\begin{split}
\varphi_p(s_1,s_2)&=-\frac{4s_2^4+4s_1^2+1}{8s_1s_2}\\
        \varphi_{t_u}(s_1,s_2)&=\frac{1}{s_2}\\
    \varphi_{t_v}(s_1,s_2)&=-\frac{4s_2^4-4s_1^2+1}{8s_1s_2}.
\end{split}
\label{eq:1+u+vParam}
\end{align}
Combining (\ref{eq:1+u+vParam}) with (\ref{eq:final3RootsParam}), we find the final change of variables for $u$ and $v$ that rationalizes all of the four square roots in (\ref{eq:SetOfSimultaneousRoots}) simultaneously:
\begin{align}
\begin{split}
        \Phi_u(s_1,s_2):&=\tilde{\varphi}_u(\varphi_{t_u}(s_1,s_2))\\
    \Phi_v(s_1,s_2):&=\tilde{\varphi}_v(\varphi_{t_v}(s_1,s_2)).
\end{split}
\end{align}
Indeed, we check that
\begin{align}
\begin{split}
        \sqrt{\Phi_u(s_1,s_2)+1}&=\frac{2s_2^2+1}{2s_2}\\
    \sqrt{\Phi_u(s_1,s_2)-1}&=\frac{2s_2^2-1}{2s_2}\\
    \sqrt{\Phi_v(s_1,s_2)+1}&=\frac{4s_2^4-4s_1^2+1}{8s_1s_2}\\
    \sqrt{\Phi_u(s_1,s_2)+\Phi_v(s_1,s_2)+1}&=\frac{4s_2^4+4s_1^2+1}{8s_1s_2}.
\end{split}
\end{align}
\label{ex:rationalizingMultipleRoots}
\end{example}

\section{Summary}
\label{sec:summary}

In this paper, we considered the problem of rationalizing roots.
This problem occurs in Feynman integral computations.
Methods from algebraic geometry allow us to address this problem.
We discussed an algorithm, which allows the rationalization of roots, whenever the associated hypersurface is irreducible and has a point of multiplicity $(d-1)$, where $d$ is the degree of the polynomial defining the hypersurface.
This algorithm covers many cases from high energy physics that admit a rational parametrization.
Not all roots are rationalizable.
In particular, this is true, if the associated hypersurface is a smooth elliptic curve or a K3 surface.
In an appendix, we discussed examples for both cases.
Our proofs for the impossibility of rationalizing the roots associated with these particular hypersurfaces may serve as a template for answering rationalization questions in similar circumstances.
We expect our results to be useful for Feynman integral computations.

\section*{Acknowledgments}
\label{sec:acknowledgments}
We want to thank Claude Duhr, Lorenzo Tancredi, Johannes Henn, Simone Zoia, Ekta Chaubey and Pascal Wasser for helpful comments. We also want to thank Dino Festi for hours of fruitful discussions.


\appendix

\section{Limitations of the algorithm}
\label{sec:limitationsOfTheAlgorithm}
There are roots, which are rationalizable, but where the associated algebraic hypersurface
does not possess a point of multiplicity $(d-1)$.
These roots cannot be rationalized with our algorithm.
We are not aware of an example from particle physics, where this is the case. 
Instead, we give an example from mathematics with no relation to physics.
Consider the root
\begin{equation}
    \sqrt{\frac{-2x^2+\sqrt{8x^2+1}-1}{2}}.
    \label{eq:lemniscateSolution}
\end{equation}
The associated curve is defined by the polynomial
\begin{equation}
    f(x,y)=x^4+2x^2y^2-x^2+y^4+y^2.
\end{equation}
Although this curve does not possess a point of multiplicity $3$, it does have a rational parametrization. 
One easily checks that replacing $x$ by 
\begin{equation}
    \varphi_x(t)=\frac{t\left(t^2+1\right)}{t^4+1}
\end{equation}
rationalizes the root given in (\ref{eq:lemniscateSolution}).
To find this change of variables, one needs to parametrize the associated curve by a family of circles instead of parametrizing it by a family of lines.
However, this algorithm is a bit more involved and a possible topic for another publication.

\section{Roots without rational parametrizations}

There are roots, which we can prove to be not rationalizable.
In this appendix, we give a few examples from particle physics.

\subsection{Roots associated with elliptic curves}
\label{sec:rootsAssociatedToEllipticCurves}

A square root of a cubic or quartic polynomial, where all zeros of the polynomial are distinct, defines (together with a rational point) a non-degenerate elliptic curve.
Roots of this type cannot be rationalized.
The obvious examples from physics are Feynman integrals where such roots occur explicitly \cite{Laporta:2004rb,MullerStach:2011ru,Adams:2013kgc,Bloch:2013tra,Adams:2014vja,Adams:2015gva,Adams:2015ydq,Sogaard:2014jla,Bloch:2016izu,Remiddi:2016gno,Adams:2016xah,Bonciani:2016qxi,vonManteuffel:2017hms,Adams:2017ejb,Bogner:2017vim,Ablinger:2017bjx,Remiddi:2017har,Bourjaily:2017bsb,Hidding:2017jkk,Broedel:2017kkb,Broedel:2017siw,Broedel:2018iwv,Adams:2018yfj,Adams:2018bsn,Adams:2018kez}. The most prominent example is given by the two-loop sunrise integral with non-vanishing internal masses.
Let us discuss here a less obvious example.
We consider the phase space integration of the double-real / single-virtual contribution to
the differential $\mathrm{N}^3\mathrm{LO}$ cross section for Higgs production in the heavy top mass limit \cite{Dulat:talk2018, Anastasiou:2015yha}.
In this part of the calculation one encounters the algebraic alphabet
\bq
 {\mathcal A}
 & = &
 \left\{
 z, 1-z, 1+z,
 1 + \sqrt{z},
 1 + \sqrt{1+4z},
 2 - z + \sqrt{z \left(z-4\right)}
 \right\}.
\eq
Following example \ref{ex:rationalizingMultipleRoots}, one easily rationalizes the fourth and the fifth letter of $\mathcal{A}$ with a single parametrization.
However, it is not possible to rationalize all letters of $\mathcal{A}$ simultaneously.
It is not even possible to rationalize the last two letters simultaneously.
We show this by contradiction.\\
Suppose there exists a rational function $\varphi_z(t)\in \mathbb{Q}(t)$ such that the last two letters of $\mathcal{A}$ are rationalized simultaneously, i.e., it is true that
\begin{equation}
    \left(1 + \sqrt{1+4\varphi_z(t)}\right)\in \mathbb{Q}(t)\hspace{8pt}\text{and}\hspace{8pt}\left(2 - \varphi_z(t) + \sqrt{\varphi_z(t) \left(\varphi_z(t)-4\right)}\right)\in \mathbb{Q}(t).
\end{equation} 
It follows that
\begin{equation}
    \sqrt{1+4\varphi_z(t)}\in \mathbb{Q}(t)\hspace{8pt}\text{and}\hspace{8pt}\sqrt{\varphi_z(t)\cdot\left(\varphi_z(t)-4\right)}\in \mathbb{Q}(t).
\end{equation} 
One concludes
\begin{equation}
    \sqrt{\left(1+4\varphi_z(t)\right)\cdot\left(\varphi_z(t)\cdot\left(\varphi_z(t)-4\right)\right)}\in \mathbb{Q}(t).
    \label{eq:productRoot}
\end{equation} 
We give the function in (\ref{eq:productRoot}) the name $\varphi_u(t)$. This means, we have found rational functions $\varphi_u(t), \varphi_z(t)\in \mathbb{Q}(t)$ which solve the polynomial equation
\begin{equation}
    u^2=(1+4z)\cdot(z(z-4)).
\label{eq:ellipticCurveViaLetters}
\end{equation}
Thus, we have found rational functions, which parametrize the algebraic curve defined by this polynomial. However, equation (\ref{eq:ellipticCurveViaLetters}) defines a smooth elliptic curve, i.e., a curve of genus $1$, and it is known for more than 150 years \cite{MR1579280} that algebraic curves can be parametrized by rational functions if and only if their genus is $0$. Contradiction.\\
We conclude that there is no rational function $\varphi_z(t)\in \mathbb{Q}(t)$ such that the last two letters of $\mathcal{A}$ are rationalized simultaneously.

\subsection{Roots associated with K3 surfaces}

Computing the master integrals for massive two-loop Bhabha scattering in QED up to order four in dimensional regularization, one can express the results almost exclusively in terms of multiple polylogarithms \cite{Henn:2013woa}. However, for one of the master integrals, it is not at all clear how to reexpress the integral in terms of these well-studied functions. The usual approach to tackle this kind of problem is to rationalize the arguments of the logarithmic forms which appear in the differential equation for the integral via an appropriate change of variables. 
This strategy is supported by the observation that rational arguments of the logarithmic forms always lead to results in terms of multiple polylogarithms.

We start with a short argument that K3 surfaces are not parametrizable by rational functions.
It is well-known that on $\mathbb{P}^1$ there does not exist a non-zero $1$-form which is holomorphic everywhere:
in fact, the $1$-form $dx$ has a pole of order $2$ at infinity, which can be seen by considering $t=1/x$, where $t$ is a local coordinate in a
neighborhood of infinity.
We can use this basic fact to prove that K3 surfaces are not parametrizable by rational functions.\\
If $X$ is a K3 surface, it comes with a natural non-zero holomorphic $2$-form $\omega$ on $X$. 
For example, if $X$ is given by a polynomial equation $f=0$, then $\omega$ can be written explicitly in local affine coordinates as the residue 
\begin{equation}
    \omega=\text{Res}\left(\frac{dy_1\wedge dy_2\wedge dy_3}{f}\right).
\end{equation}
Now, if the K3 surface $X$ would be parametrizable by rational functions, we would have a rational map $\varphi:\mathbb{P}^2\dashrightarrow X$.
We define $\Sigma$ to be the finite set of points in $\mathbb{P}^2$ for which $\varphi$ is not defined. 
The pullback $\varphi^*\omega$ of $\omega$ by $\varphi$ is a holomorphic $2$-form on $\mathbb{P}^2\backslash\Sigma$ which is locally given by
$\varphi^*\omega=A(y_1,y_2)\hspace{2pt}dy_1\wedge dy_2$, where $A(y_1,y_2)$ is a holomorphic function in $y_1$ and $y_2$.
But such a $2$-form cannot exist, as can be seen as follows: pick a line $l\subset\mathbb{P}^2$ which does not pass through any of the points of
$\Sigma$ and on which $A(y_1,y_2)$ does not vanish identically.
Note that for every line $l$ in $\mathbb{P}^2$ we have $l\simeq\mathbb{P}^1$.
Without loss of generality, i.e., by an appropriate change of coordinates, we can assume that this line is given by $l: y_1=0$.
In this case, a normal vector field for $l$ is given by $\vartheta=\partial_{y_1}$.
But then, the contraction $\varphi^*\omega\intprod\vartheta=A(0,y_2)\hspace{1pt}dy_2$ of the holomorphic $2$-form $\varphi^*\omega$ with the normal
vector field $\vartheta$ is a non-zero $1$-form which is holomorphic everywhere on $l\simeq\mathbb{P}^1$.
Contradiction.

Let us now return to two-loop Bhabha scattering.
The integral under consideration is $f_{11}^{(4)}$ of \cite{Henn:2013woa} and has the following differential equation:

\begin{align}
\begin{split}
            df_{11}^{(4)}=&
g_{1}d\log\left(\frac{1-Q}{1+Q}\right)\\
+&g_{2}d\log\left(\frac{(1+x)+(1-x)Q}{(1+x)-(1-x)Q}\right)\\
+&g_{3}d\log\left(\frac{(1+y)+(1-y)Q}{(1+y)-(1-y)Q}\right).
\end{split}
\end{align}

The most promising strategy in order to show that $f_{11}^{(4)}$ evaluates to multiple polylogarithms, is to find a change of variables $\varphi_x(t_1,t_2)$ and $\varphi_y(t_1,t_2)$ such that the arguments of the above logarithms become rational. This is, for instance, the case when the square root

\begin{equation}
    Q=\sqrt{\frac{(x+y)(1+xy)}{x+y-4xy+x^2y+xy^2}}
    \label{eq:defOfK3}
\end{equation}

is rationalized. However, although being the most promising approach to the problem, we will show below that $Q$ itself cannot be rationalized. A standard reference for this subject is \cite{MR2030225}.\\
Let us give a brief sketch of the proof: by defining
\begin{equation}
    u:=\frac{x+y}{Q},
\end{equation}
squaring (\ref{eq:defOfK3}) and clearing denominators, we obtain the following quartic algebraic surface:
\begin{equation}
    V: u^2\cdot(1 + x y)=(x + y)\cdot\left(x + y - 4 x y + x^2 y + x y^2\right).
\end{equation}
In order to show that the square root under consideration is not rationalizable, we need to check that all singularities of the quartic surface $V$ are of multiplicity 2, showing that $V$ has singularities of ADE type only. First of all, we consider the projective closure $\tilde{V}$, which is defined by the homogeneous polynomial 
\begin{equation}   
F(x,y,u,z)=u^2\cdot\left(z^2 + x y\right)-(x+y)\cdot\left((x + y)z^2 - 4 x yz + x^2 y + x y^2\right).
\end{equation}
The singular points $[x:y:u:z]\in \Sigma\subset\tilde{V}$ of this projective hypersurface are easily computed. One obtains 
\begin{align}
\begin{split}
     \Sigma=\{&[1:1:0:1],[1:-1:0:0],[0:1:1:0],[0:1:-1:0],\\
&[0:0:1:0],[1:0:1:0],[1:0:-1:0],[0:0:0:1]\}.
\end{split}
\end{align}
Checking the partial derivatives at these points, we see that at each point, at least one of the second derivatives is non-zero, i.e., all singular points of $\tilde{V}$ are of multiplicity 2. We conclude that all singularities are of ADE type. Using $\mathtt{SINGULAR}$ \cite{DGPS, GLP, PDSL}, we find the following classification:
\begin{center}
    \begin{tabular}{ | c | c |}
    \hline
    Singularity & Type \\ \hline\hline
    $[1:1:0:1], [1:-1:0:1], [0:0:1:0]$ & $A_1$ \\ \hline
    $[0:1:1:0], [0:1:-1:0], [1:0:1:0], [1:0:-1:0]$ & $A_2$ \\ \hline
    $[0:0:0:1]$ & $A_3$ \\
    \hline
    \end{tabular}
\end{center}
Consequently, the hypersurface associated with the square root under consideration is a K3 surface by the above arguments and in turn not rationalizable.\\
Notice that this statement is just about $Q$ itself. It does not prove that $f_{11}^{(4)}$ cannot be written in terms of multiple polylogarithms. However, it is a very strong indication that it is not possible.\\
A detailed mathematical analysis of the above K3 is carried out in \cite{Festi:082018,Festi:092018}.
We also want to mention that there is recent progress in rewriting $f_{11}^{(4)}$ in terms of known functions. In \cite{Duhr:2018}, it will be shown that $f_{11}^{(4)}$ can be expressed in terms of elliptic polylogarithms. This can be achieved by exploiting the fact that the above K3 has an elliptic fibration.
\par\vspace{\baselineskip}
We may also utilize K3 surfaces to prove statements similar to the one in appendix \ref{sec:rootsAssociatedToEllipticCurves}.
From \cite{Becchetti:2017abb} and example \ref{ex:rationalizingMultipleRoots} we know that all roots of the set
\begin{equation}
    \mathcal{A}=\left\{\sqrt{u+1},\sqrt{u-1},\sqrt{v+1},\sqrt{u+v+1}\right\}
\end{equation}
can be rationalized simultaneously. 
Let us now consider the case, where $\mathcal{A}$ has one additional root $\sqrt{16u+(4+v)^2}$.
The choice of this new alphabet
\begin{equation}
 \mathcal{A}^\prime=\left\{\sqrt{u+1}, \sqrt{u-1}, \sqrt{v+1}, \sqrt{u+v+1}, \sqrt{16u+(4+v)^2}\right\}
\end{equation}
corresponds to the alphabet of topology A in \cite{Becchetti:2017abb}. Using the fact that K3 surfaces are not parametrizable by rational functions, we are able to prove that the five letters of $\mathcal{A}^\prime$ can not be rationalized simultaneously.
Analogously to appendix \ref{sec:rootsAssociatedToEllipticCurves}, we will show this by contradiction.\\
Suppose there exist rational functions $\varphi_u(t_1,t_2),\varphi_v(t_1,t_2)\in \mathbb{Q}(t_1,t_2)$ such that all letters of $\mathcal{A}^\prime$ are rationalized simultaneously, i.e., it is true that
\begin{align}
\begin{split}
    &\sqrt{\varphi_u(t_1,t_2)+1}\in \mathbb{Q}(t_1,t_2),\hspace{8pt}\sqrt{\varphi_u(t_1,t_2)-1}\in \mathbb{Q}(t_1,t_2),\\
    &\sqrt{\varphi_v(t_1,t_2)+1}\in \mathbb{Q}(t_1,t_2),\hspace{8pt}\sqrt{\varphi_u(t_1,t_2)+\varphi_v(t_1,t_2)+1}\in \mathbb{Q}(t_1,t_2)
\end{split}
\end{align} 
and
\begin{equation}
    \sqrt{16\varphi_u(t_1,t_2)+(4+\varphi_v(t_1,t_2))^2}\in \mathbb{Q}(t_1,t_2).
\end{equation}
It follows that
\begin{equation}
    \sqrt{\left(\varphi_u+1\right)\cdot\left(\varphi_u-1\right)\cdot\left(\varphi_v+1\right)\cdot\left(\varphi_u+\varphi_v+1\right)\cdot\left(16\varphi_u+(4+\varphi_v)^2\right)}\in \mathbb{Q}(t_1,t_2),
\label{eq:productK3Root}
\end{equation}
where $\varphi_u\equiv\varphi_u(t_1,t_2)$, $\varphi_v\equiv\varphi_v(t_1,t_2)$.
We give the function in (\ref{eq:productK3Root}) the name $\varphi_w(t_1,t_2)$. This means, we have found rational functions $\varphi_u(t_1,t_2), \varphi_v(t_1,t_2), \varphi_w(t_1,t_2)\in \mathbb{Q}(t_1,t_2)$ which solve the polynomial equation 
\begin{equation}
    w^2=\left(u+1\right)\cdot\left(u-1\right)\cdot\left(v+1\right)\cdot\left(u+v+1\right)\cdot\left(16u+(4+v)^2\right).
\label{eq:K3ViaLetters}
\end{equation}
But this means that we have found rational functions, which parametrize the algebraic surface defined by this polynomial.
However, equation (\ref{eq:K3ViaLetters}) defines a K3 surface, which can be seen as follows:
homogenizing the right-hand side, we can write the hypersurface defined by (\ref{eq:K3ViaLetters}) as
\begin{equation}
    X: w^2=F_6(u,v,z),
\end{equation}
which defines a hypersurface of degree $6$ in the weighted projective space $\mathbb{P}(1,1,1,3)$, where $u,v$ and $z$ are homogeneous coordinates of weight $1$, $w$ is a homogeneous coordinate of weight $3$ and $F_6$ is a homogeneous polynomial of degree $6$. (For an introduction to weighted projective space, see \cite{2016arXiv160402441H}.)
This corresponds to a double cover $\pi: X\rightarrow\mathbb{P}^2$ ramified along a sextic curve $C\subset\mathbb{P}^2$ with defining polynomial $F_6$.
Since the singularities of a double cover are always inherited from its ramification locus, it suffices to study the singularities of the curve $C: F_6=0$.
To show that (\ref{eq:K3ViaLetters}) defines a K3, it is therefore enough to show that all singularities of $C$ are of ADE type.
Notice, however, that we make explicit use of the fact that we are dealing with a sextic ramification locus. 
A quartic double cover, for example, would allow for a rational parametrization.\\
The curve $C$ is given by a union of four lines $L_i: l_i=0$, $i=1,2,3,4$, together with a smooth conic $Q^\prime: q=0$
\begin{equation}
    C: l_1\cdot l_2\cdot l_3\cdot l_4\cdot q=0
\end{equation}
with polynomials
\begin{equation}
    l_1(u,v,z)=u+z,\hspace{8pt}l_2(u,v,z)=u-z,\hspace{8pt}l_3(u,v,z)=v+z,\hspace{8pt}l_4(u,v,z)=u+v+z
\end{equation}
and
\begin{equation}
    q(u,v,z)=16uz+(4z+v)^2.
\end{equation}
Since all components of $C$ define smooth curves themselves, possible singularities of $C$ can only arise from intersection points of these components.
The four lines intersect in six points:
\begin{align}
\begin{split}
    L_1\cap L_2&=[0:1:0],\hspace{8pt}L_1\cap L_3=[-1:-1:1],\hspace{8pt}L_1\cap L_4=[-1:0:1],\\
    L_2\cap L_3&=[1:-1:1],\hspace{8pt}L_2\cap L_4=[1:-2:1],\hspace{8pt}L_3\cap L_4=[0:-1:1].
\end{split}
\end{align}
Five of these points are $A_1$ singularities of $C$, as they are just a simple intersection point of two lines.
However, $[-1:0:1]$ is an exception.
The reason is that $[-1:0:1]$ is a point of $Q^\prime$, as well. 
So instead of two smooth branches, we see that actually, three smooth branches of $C$ pass through this point. 
Consequently, $[-1:0:1]$ defines a $D_4$ singularity.
Calculating the intersection points of $Q^\prime$ with each line, we obtain
\begin{align}
    \begin{split}
    L_1\cap Q^\prime&=\left\{[-1:0:1], [-1:-8:1]\right\},\hspace{8pt}L_2\cap Q^\prime=\left\{[1:-4-4i:1], [1:-4+4i:1]\right\},\\
    L_3\cap Q^\prime&=\left\{[1:0:0], \left[-9/16:-1:1\right]\right\},\hspace{8pt}L_4\cap Q^\prime=\left\{[-1:0:1], [-9:8:1]\right\}.
\end{split}
\end{align}
Despite $[-1:0:1]$, all of these points are again intersections of two smooth branches of $C$.
Therefore, the singular locus of $C$ is given by eleven $A_1$ and a single $D_4$ singularity.
We see that all singularities of $C$ and in turn all singularities of the hypersurface (\ref{eq:K3ViaLetters}) are of ADE type.
It follows that (\ref{eq:K3ViaLetters}) defines a K3 surface and is in turn not parametrizable by rational functions. Contradiction.\\
We conclude that there are no rational functions $\varphi_u(t_1,t_2), \varphi_v(t_1,t_2)\in \mathbb{Q}(t_1,t_2)$ such that all letters of $\mathcal{A}^\prime$ are rationalized simultaneously.

\subsection{Non-rationalizability of alphabets and Kummer coverings}

Let us put the ideas of the two preceding subsections into a slightly more general context.
Suppose we want to rationalize an alphabet containing multiple roots, e.g.,
\begin{equation}
    \left\{\sqrt{Q_1(x_1,\ldots,x_n)},\sqrt{Q_2(x_1,\ldots,x_n)},\ldots,\sqrt{Q_r(x_1,\ldots,x_n)}\right\}.
\end{equation}
If we found such a parametrization, this would provide us with a rational parametrization of the Kummer covering
\begin{equation}
    w_1^2=Q_1,\hspace{4pt}w_2^2=Q_2,\hspace{4pt}\ldots\hspace{4pt},\hspace{4pt}w_r^2=Q_r.
\end{equation}
These equations map to the hypersurface
\begin{equation}
    w^2=Q_1\cdot Q_2\cdots Q_r
\label{eq:kummerHypersurface}
\end{equation}
via
\begin{equation}
    (w_1,\ldots,w_r,x_1,\ldots,x_n)\mapsto(w_1w_2\cdots w_r,x_1,\ldots,x_n).
\end{equation}
If the Kummer covering is parametrizable, then the hypersurface (\ref{eq:kummerHypersurface}) is parametrizable, as well. 
To show non-rationalizability of the alphabet, it is therefore sufficient to show that the hypersurface (\ref{eq:kummerHypersurface}) does not possess a rational parametrization.
Let us stress that the converse statement is not true: the rationality of the above hypersurface does not necessarily imply rationality of the corresponding Kummer covering.\\
Actually, (\ref{eq:kummerHypersurface}) is not the only hypersurface that one can associate with the Kummer covering.
In some cases, it might be easier to prove non-rationalizability of a different hypersurface, e.g.,
\begin{equation}
    Q_1\cdot w^2=Q_2\cdot Q_3\cdots Q_r.
\label{eq:kummerHypersurface2}
\end{equation}
The Kummer covering maps to this hypersurface via
\begin{equation}
    (w_1,\ldots,w_r,x_1,\ldots,x_n)\mapsto\left(\frac{w_2\cdots w_r}{w_1},x_1,\ldots,x_n\right).
\end{equation}
Notice that, although being different hypersurfaces, (\ref{eq:kummerHypersurface}) and (\ref{eq:kummerHypersurface2}) are, however, birational and thus either rationalizable at the same time or not rationalizable at the same time.

\section{A theorem on a particular type of square roots}

In this appendix, we will show that affine hypersurfaces of dimension $n$ and \textit{even} degree $d$ with defining equation
\begin{equation}
    u^2=\left(F_{\frac{d}{2}}(x_1,\ldots,x_n,1)\right)^2-4\cdot F_{\frac{d}{2}+1}(x_1,\ldots,x_n,1)\cdot F_{\frac{d}{2}-1}(x_1,\ldots,x_n,1)
\label{eq:1steq}
\end{equation}
are rational, if and only if the hypersurface defined by
\begin{equation}
    F_{\frac{d}{2}+1}(x_1,\ldots,x_n,z)+F_{\frac{d}{2}}(x_1,\ldots,x_n,z)+F_{\frac{d}{2}-1}(x_1,\ldots,x_n,z)=0
\label{eq:2ndeq}
\end{equation}
is rational, where $F_k\in \mathbb{C}[x_1,\ldots,x_n,z]$ are homogeneous polynomials in $n+1$ variables of degree $k$.
The proof of the statement is constructive.
Given a rational parametrization of one of the above hypersurfaces, we will provide a prescription how to obtain a rational parametrization of the respective other.
\par\vspace{\baselineskip}
Peering at the very special form of (\ref{eq:1steq}) and (\ref{eq:2ndeq}), one might think that the whole scenario is too special and constrained to have any relevance for practical applications. However, quite recently it turned out \cite{Bourjaily:2018aeq} that precisely square roots of type
\begin{equation}
    \sqrt{\left(F_{\frac{d}{2}}\right)^2-4\cdot F_{\frac{d}{2}+1}\cdot F_{\frac{d}{2}-1}}
\label{eq:squareroot}
\end{equation}
and their rational parametrizations are crucial for direct Feynman-parametric loop integration of a large class of planar multi-loop integrals.
For instance, let us consider the following root appearing in \cite{Bourjaily:2018aeq}:
\begin{equation}
    \Delta_7^{\{123567\}}=\sqrt{(1-u_1-u_2-u_3+u_2u_3u_4)^2-4u_1u_2u_3\cdot(1-u_4)}.
\label{eq:delta7}
\end{equation}
Obviously, one has $d=6$ in this case. Homogenizing the equation, it is easy to see that we can write
\begin{equation}
    \Delta_7^{\{123567\}}=\sqrt{\left(F_3(u_1,\ldots,u_4,1)\right)^2-4\cdot F_4(u_1,\ldots,u_4,1)\cdot F_2(u_1,\ldots,u_4,1)}
\end{equation}
with the choice
\begin{align}
\begin{split}
    F_2(u_1,\ldots,u_4,z)&=u_3\cdot(z-u_4)\\
    F_3(u_1,\ldots,u_4,z)&=z^3-z^2\cdot(u_1+u_2+u_3)+u_2u_3u_4\\
    F_4(u_1,\ldots,u_4,z)&=z^2u_1u_2.
\end{split}
\end{align}
Now, the point is that, using the theorem of this appendix, we can reformulate the question and, instead of asking for a rational parametrization of 
\begin{equation}
    \left(\Delta_7^{\{123567\}}\right)^2=\left(F_3(u_1,\ldots,u_4,1)\right)^2-4\cdot F_4(u_1,\ldots,u_4,1)\cdot F_2(u_1,\ldots,u_4,1),
\end{equation}
we can try to find a rational parametrization of
\begin{equation}
    0=F_4(u_1,\ldots,u_4,z)+F_3(u_1,\ldots,u_4,z)+F_2(u_1,\ldots,u_4,z).
\label{eq:delta7transform}
\end{equation}
and utilize the parametrization of the latter hypersurface to rationalize the original square root $\Delta_7^{\{123567\}}$.
Viewing the square root from this perspective, we only have to determine a point of multiplicity $3$ instead of a point of multiplicity $5$. Using definition \ref{def:Multiplicity}, we need to solve
\begin{equation}
    \sum_{k=0}^{d-2}\begin{pmatrix}
    n+k-1 \\
    k 
\end{pmatrix}
\end{equation}
equations in order to find a point of multiplicity $(d-1)$. 
So to determine a point of multiplicity $5$ of the projective hypersurface corresponding to (\ref{eq:delta7}), we would have to solve $126$ equations.
Using our theorem and trying to find a point of multiplicity $3$ for the projective closure of (\ref{eq:delta7transform}) instead, we only need to solve $28$ equations, which is already a huge improvement and becomes even more significant when we consider roots with arguments of higher degree and a higher number of variables.\\
Solving these $28$ equations, we find that the hypersurface (\ref{eq:delta7transform}) has four points of multiplicity $3$ at infinity:
\begin{equation}
    \{[1:0:1:0:0:0],[1:0:0:1:0:0],[0:1:0:0:0:0],[1:0:0:0:0:0]\}.
\end{equation}
So in order to rationalize $\Delta_7^{\{123567\}}$, we simply pick one of these points, apply the main algorithm of this paper to rationalize (\ref{eq:delta7transform}) and finally use the theorem presented below to transform the parametrization of (\ref{eq:delta7transform}) into a parametrization for $\Delta_7^{\{123567\}}$.
\par\vspace{\baselineskip}
Let us begin by proving that rationality of the algebraic hypersurface defined by (\ref{eq:1steq}) implies rationality of the algebraic hypersurface defined by (\ref{eq:2ndeq}).
\begin{theorem}
    We consider a rational affine complex algebraic hypersurface $V\subset\mathbb{A}^{n+1}(\mathbb{C})$ of dimension $n$ defined by a polynomial equation of the form
\begin{equation}
    u^2=f_d(x_1,\ldots,x_n),
\end{equation}
where $f_d(x_1,\ldots,x_n)$ is a polynomial in $n$ variables of even degree $d$. Furthermore, we assume that $f_d(x_1,\ldots,x_n)$ can be written as
\begin{equation}
    f_d(x_1,\ldots,x_n)=\left(F_{\frac{d}{2}}(x_1,\ldots,x_n,1)\right)^2-4\cdot F_{\frac{d}{2}+1}(x_1,\ldots,x_n,1)\cdot F_{\frac{d}{2}-1}(x_1,\ldots,x_n,1),
\end{equation}
where $F_{\frac{d}{2}}(x_1,\ldots,x_n,z)$, $F_{\frac{d}{2}+1}(x_1,\ldots,x_n,z)$ and $F_{\frac{d}{2}-1}(x_1,\ldots,x_n,z)$ are homogeneous polynomials in $n+1$ variables of degree $\frac{d}{2}$, $\frac{d}{2}+1$ and $\frac{d}{2}-1$, respectively.
If
\begin{equation}
    \left(\varphi_u^V(t_1,\ldots,t_n),\varphi^V_{x_1}(t_1,\ldots,t_n),\ldots,\varphi^V_{x_n}(t_1,\ldots,t_n)\right)
\end{equation}
is a rational parametrization of $V$, then one can determine a rational parametrization
\begin{equation}
    \left(\varphi^W_{x_1}(t_1,\ldots,t_n),\ldots,\varphi^W_{x_n}(t_1,\ldots,t_n),\varphi^W_z(t_1,\ldots,t_n)\right)
\end{equation}
of the complex affine algebraic hypersurface $W\subset\mathbb{A}^{n+1}(\mathbb{C})$ of dimension $n$ and degree $\frac{d}{2}+1$, which is defined by the equation
\begin{equation}
    F_{\frac{d}{2}+1}(x_1,\ldots,x_n,z)+F_{\frac{d}{2}}(x_1,\ldots,x_n,z)+F_{\frac{d}{2}-1}(x_1,\ldots,x_n,z)=0.
\label{eq:sum}
\end{equation}
\label{thm:maintheorem}
\end{theorem}
\textbf{Proof:}\\
We start with the formal ansatz
\begin{align}
\begin{split}
       \varphi^W_{x_1} &=\lambda\cdot x^\prime_1\\
&\vdots\\
\varphi^W_{x_n} &=\lambda\cdot x^\prime_n\\
\varphi^W_z &=\lambda.
\end{split}
\label{al:xiprimes}
\end{align}
Plugging this ansatz into equation (\ref{eq:sum}), we obtain
\begin{equation}
    \lambda^{\frac{d}{2}+1}\cdot F_{\frac{d}{2}+1}(x^\prime_1,\ldots,x^\prime_n,1)+\lambda^{\frac{d}{2}}\cdot F_{\frac{d}{2}}(x^\prime_1,\ldots,x^\prime_n,1)+\lambda^{\frac{d}{2}-1}\cdot F_{\frac{d}{2}-1}(x^\prime_1,\ldots,x^\prime_n,1)=0.
\end{equation}
Despite the solution $\lambda_0=0$, there are two other solutions $\lambda_{\pm}$. For example, $\lambda_+$ is given by
\begin{align}
\begin{split}
    \lambda_+&=-\frac{F_{\frac{d}{2}}(x^\prime_1,\ldots,x^\prime_n,1)}{2\cdot F_{\frac{d}{2}+1}(x^\prime_1,\ldots,x^\prime_n,1)}+\sqrt{\left(\frac{F_{\frac{d}{2}}(x^\prime_1,\ldots,x^\prime_n,1)}{2\cdot F_{\frac{d}{2}+1}(x^\prime_1,\ldots,x^\prime_n,1)}\right)^2-\frac{F_{\frac{d}{2}-1}(x^\prime_1,\ldots,x^\prime_n,1)}{F_{\frac{d}{2}+1}(x^\prime_1,\ldots,x^\prime_n,1)}}\\
&=-\frac{F_{\frac{d}{2}}(x^\prime_1,\ldots,x^\prime_n,1)}{2\cdot F_{\frac{d}{2}+1}(x^\prime_1,\ldots,x^\prime_n,1)}+\frac{u}{2\cdot F_{\frac{d}{2}+1}(x^\prime_1,\ldots,x^\prime_n,1)},
\end{split}
\end{align}
where we defined $u$ to be
\begin{equation}
    u:= \sqrt{\left(F_{\frac{d}{2}}(x^\prime_1,\ldots,x^\prime_n,1)\right)^2-4\cdot F_{\frac{d}{2}+1}(x^\prime_1,\ldots,x^\prime_n,1)\cdot F_{\frac{d}{2}-1}(x^\prime_1,\ldots,x^\prime_n,1)}.
\label{eq:u}
\end{equation}
Now, by assumption, we have a rational parametrization of the hypersurface $V$. Let us therefore substitute $(u,x_1^\prime,\ldots,x_n^\prime)$ by
\begin{equation}
    \left(\varphi_u^V(t_1,\ldots,t_n),\varphi^V_{x_1}(t_1,\ldots,t_n),\ldots,\varphi^V_{x_n}(t_1,\ldots,t_n)\right).
\label{eq:param}
\end{equation}
In particular, $\varphi^V_u(t_1,\ldots,t_n)$ provides us with a rational expression for the square root $u$ in (\ref{eq:u}). But this means that, using the given parametrization (\ref{eq:param}), we can express $\lambda_+$ as 
\begin{align}
\begin{split}
    \lambda_+(t_1,\ldots,t_n)=&-\frac{F_{\frac{d}{2}}\left(\varphi^V_{x_1}(t_1,\ldots,t_n),\ldots,\varphi^V_{x_n}(t_1,\ldots,t_n),1\right)}{2\cdot F_{\frac{d}{2}+1}\left(\varphi^V_{x_1}(t_1,\ldots,t_n),\ldots,\varphi^V_{x_n}(t_1,\ldots,t_n),1\right)}\\
&+\frac{\varphi^V_u(t_1,\ldots,t_n)}{2\cdot F_{\frac{d}{2}+1}\left(\varphi^V_{x_1}(t_1,\ldots,t_n),\ldots,\varphi^V_{x_n}(t_1,\ldots,t_n),1\right)}
\end{split}
\end{align}
turning $\lambda_+$ into a rational function of $t_1,\ldots,t_n$.\\
Since $\lambda_+(t_1,\ldots,t_n)$ as well as
$\varphi^V_{x_1}(t_1,\ldots,t_n),\ldots,\varphi^V_{x_n}(t_1,\ldots,t_n)$ are rational functions of $t_1,\ldots,t_n$ and, additionally, $\lambda_+$ was precisely chosen in a way such that (\ref{al:xiprimes}) solves the defining equation (\ref{eq:sum}) of $W$, we conclude that
\begin{align}
\begin{split}
    \varphi^W_{x_1}(t_1,\ldots,t_n)&=\lambda_+(t_1,\ldots,t_n)\cdot \varphi^V_{x_1}(t_1,\ldots,t_n)\\
&\vdots\\
\varphi^W_{x_n}(t_1,\ldots,t_n)&=\lambda_+(t_1,\ldots,t_n)\cdot \varphi^V_{x_n}(t_1,\ldots,t_n)\\
\varphi^W_z(t_1,\ldots,t_n)&=\lambda_+(t_1,\ldots,t_n)
\end{split}
\end{align}
is the sought after rational parametrization of $W$, proving $W$ to be a rational algebraic hypersurface.\hfill\(\Box\)
\par\vspace{\baselineskip}
Let us now prove the, for us more important, converse statement.
\begin{theorem}
        We consider a rational affine complex algebraic hypersurface $W\subset\mathbb{A}^{n+1}(\mathbb{C})$ of dimension $n$ and degree $\frac{d}{2}+1$, where $d\in \mathbb{N}$ is even. Suppose $W$ is defined by a polynomial equation of the form
\begin{equation}
    F_{\frac{d}{2}+1}(x_1,\ldots,x_n,z)+F_{\frac{d}{2}}(x_1,\ldots,x_n,z)+F_{\frac{d}{2}-1}(x_1,\ldots,x_n,z)=0,
\end{equation}
where $F_k$ are homogeneous polynomials of degree $k$.
If
\begin{equation}
    \left(\varphi^W_{x_1}(t_1,\ldots,t_n),\ldots,\varphi^W_{x_n}(t_1,\ldots,t_n),\varphi^W_z(t_1,\ldots,t_n)\right)
\end{equation}
is a rational parametrization of $W$, then one can determine a rational parametrization
\begin{equation}
    \left(\varphi^V_u(t_1,\ldots,t_n),\varphi^V_{x_1}(t_1,\ldots,t_n),\ldots,\varphi^V_{x_n}(t_1,\ldots,t_n)\right)
\end{equation}
of the affine hypersurface $V\subset\mathbb{A}^{n+1}(\mathbb{C})$ defined by the polynomial equation
\begin{equation}
    u^2=\left(F_{\frac{d}{2}}(x_1,\ldots,x_n,1)\right)^2-4\cdot F_{\frac{d}{2}+1}(x_1,\ldots,x_n,1)\cdot F_{\frac{d}{2}-1}(x_1,\ldots,x_n,1).
\end{equation}
\label{thm:maintheorem2}
\end{theorem}
\textbf{Proof:}\\
We assume
\begin{equation}
    \left(\varphi^W_{x_1},\ldots,\varphi^W_{x_n},\varphi^W_z\right):=\left(\varphi^W_{x_1}(t_1,\ldots,t_n),\ldots,\varphi^W_{x_n}(t_1,\ldots,t_n),\varphi^W_z(t_1,\ldots,t_n)\right)
\end{equation}
to be a rational parametrization of $W$, i.e., rational functions $\left(\varphi^W_{x_1},\ldots,\varphi^W_{x_n},\varphi^W_z\right)$ satisfying
\begin{equation}
    0=F_{\frac{d}{2}+1}\left(\varphi^W_{x_1},\ldots,\varphi^W_{x_n},\varphi^W_z\right)+F_{\frac{d}{2}}\left(\varphi^W_{x_1},\ldots,\varphi^W_{x_n},\varphi^W_z\right)+F_{\frac{d}{2}-1}\left(\varphi^W_{x_1},\ldots,\varphi^W_{x_n},\varphi^W_z\right).
\end{equation}
One can rewrite this equation like
\begin{align}
\begin{split}
        0&=\left(\varphi^W_z\right)^{\frac{d}{2}+1}\cdot F_{\frac{d}{2}+1}\left(\frac{\varphi^W_{x_1}}{\varphi^W_z},\ldots,\frac{\varphi^W_{x_n}}{\varphi^W_z},1\right)\\
&+\left(\varphi^W_z\right)^{\frac{d}{2}}\cdot F_{\frac{d}{2}}\left(\frac{\varphi^W_{x_1}}{\varphi^W_z},\ldots,\frac{\varphi^W_{x_n}}{\varphi^W_z},1\right)\\
&+\left(\varphi^W_z\right)^{\frac{d}{2}-1}\cdot F_{\frac{d}{2}-1}\left(\frac{\varphi^W_{x_1}}{\varphi^W_z},\ldots,\frac{\varphi^W_{x_n}}{\varphi^W_z},1\right).
\end{split}
\label{eq:phiwzequation}
\end{align}
By assumption, $\varphi^W_z$ is a non-zero solution of (\ref{eq:phiwzequation}), so it has to fulfill one of the following two equations
\begin{align}
\begin{split}
    \varphi^W_z&=-\frac{F_{\frac{d}{2}}\left(\frac{\varphi^W_{x_1}}{\varphi^W_z},\ldots,\frac{\varphi^W_{x_n}}{\varphi^W_z},1\right)}{2\cdot F_{\frac{d}{2}+1}\left(\frac{\varphi^W_{x_1}}{\varphi^W_z},\ldots,\frac{\varphi^W_{x_n}}{\varphi^W_z},1\right)}\\
&\pm \sqrt{\left(\frac{F_{\frac{d}{2}}\left(\frac{\varphi^W_{x_1}}{\varphi^W_z},\ldots,\frac{\varphi^W_{x_n}}{\varphi^W_z},1\right)}{2\cdot F_{\frac{d}{2}+1}\left(\frac{\varphi^W_{x_1}}{\varphi^W_z},\ldots,\frac{\varphi^W_{x_n}}{\varphi^W_z},1\right)}\right)^2-4\cdot \frac{F_{\frac{d}{2}-1}\left(\frac{\varphi^W_{x_1}}{\varphi^W_z},\ldots,\frac{\varphi^W_{x_n}}{\varphi^W_z},1\right)}{F_{\frac{d}{2}+1}\left(\frac{\varphi^W_{x_1}}{\varphi^W_z},\ldots,\frac{\varphi^W_{x_n}}{\varphi^W_z},1\right)}}.
\end{split}
\label{eq:2ndpq}
\end{align}
Rearranging these equations and taking the square, we get
\begin{equation}
    \left(\varphi^V_u\right)^2=\left(F_{\frac{d}{2}}\left(\varphi^V_{x_1},\ldots,\varphi^V_{x_n},1\right)\right)^2-4\cdot F_{\frac{d}{2}+1}\left(\varphi^V_{x_1},\ldots,\varphi^V_{x_n},1\right)\cdot F_{\frac{d}{2}-1}\left(\varphi^V_{x_1},\ldots,\varphi^V_{x_n},1\right),
\label{eq:endveq}
\end{equation}
where we defined
\begin{align}
\begin{split}
 \varphi^V_u&:=2\cdot F_{\frac{d}{2}+1}\left(\frac{\varphi^W_{x_1}}{\varphi^W_z},\ldots,\frac{\varphi^W_{x_n}}{\varphi^W_z},1\right)\cdot \varphi^W_z+F_{\frac{d}{2}}\left(\frac{\varphi^W_{x_1}}{\varphi^W_z},\ldots,\frac{\varphi^W_{x_n}}{\varphi^W_z},1\right)\\
\varphi^V_{x_1}&:=\frac{\varphi^W_{x_1}}{\varphi^W_z}\\
&\vdots\\
\varphi^V_{x_n}&:=\frac{\varphi^W_{x_n}}{\varphi^W_z}.   
\end{split}
\end{align}
We conclude that these definitions provide the sought after rational parametrization of $V$. \hfill\(\Box\)


\bibliography{Bibliography}
\bibliographystyle{/home/stefanw/latex-style/JHEP_new2}


\end{document}